\documentclass{elsarticle}
\usepackage{hyperref}
\usepackage[separate-uncertainty=true]{siunitx}
\usepackage{amssymb}
\usepackage{epsfig}
\usepackage{pifont}
\usepackage{ulem}
\usepackage{subcaption}
\usepackage{graphicx}
\begin{document}
	
	\begin{frontmatter}
		
		\title{Inter-Calibration of Atmospheric Cherenkov Telescopes with UAV-based Airborne Calibration System.}
		
		\author{Anthony M. Brown}
		\address{Centre for Advanced Instrumentation, Department of Physics, University of Durham, South Road, Durham, DH1 3LE, UK}
		\ead{anthony.brown@durham.ac.uk}
		
		\author{Jacques Muller}
		\address{Laboratoire Leprince-Ringuet, CNRS/IN2P3, Ecole polytechnique, Institut Polytechnique de Paris, Palaiseau, France}
		\ead{jacques.muller@polytechnique.edu}
		
		\author{Mathieu de Naurois}
		\address{Laboratoire Leprince-Ringuet, CNRS/IN2P3, Ecole polytechnique, Institut Polytechnique de Paris, Palaiseau, France}
		\ead{denauroi@in2p3.fr}
		
		\author{Paul Clark}
		\address{Centre for Advanced Instrumentation, Department of Physics, University of Durham, South Road, Durham, DH1 3LE, UK}	
		\begin{abstract}
			
		The recent advances in the flight capability of remotely piloted aerial vehicles (here after referred to as UAVs) have afforded the astronomical community the possibility of a new telescope calibration technique: UAV-based calibration. Building upon a feasibility study which characterised the potential that a UAV-based calibration system has for the future Cherenkov Telescope Array, we created a first-generation UAV-calibration prototype and undertook a field-campaign of inter-calibrating the sensitivity of the H.E.S.S. telescope array with two successful calibration flights. In this paper we report the key results of our first test campaign: firstly, by comparing the intensity of the UAV-calibration events, as recorded by the individual HESS-I cameras, we find that a UAV-based inter-calibration is consistent with the standard muon inter-calibration technique at the level of \SI{5.4}{\%} and \SI{5.8}{\%} for the two individual UAV-calibration runs. Secondly, by comparing the position of the UAV-calibration signal on the camera focal plane, for a variety of telescope pointing models, we were able to constrain the pointing accuracy of the HESS-I telescopes at the tens of arc-second accuracy level. This is consistent with the pointing accuracy derived from other pointing calibration methods. Importantly both the inter-calibration and pointing accuracy results were achieved with a first-generation UAV-calibration prototype, which eludes to the potential of the technique and highlights that a UAV-based system is a viable calibration technique for current and future ground-based $\gamma$-ray telescope arrays.  
		\end{abstract}
		
		
	\end{frontmatter}
	
	\section{Introduction}
	Advances in UAV technology have made them an attractive possibility as airborne calibration platforms for astronomical facilities (e.g. \cite{matthews2013,TA2015,Chang_2015,basden2018,2018Brown}). This is especially true for telescope arrays, where the maneuverability, flexibility and versatility of a UAV-based calibration system allows us to rapidily calibrate the numerous detector elements spread over a large area. While the use of many telescopes working as an array affords improvements in sensitivity, angular and energy resolution, it also introduces additional operational parameters, such as telescope-to-telescope variations in sensitivity, that need to be characterised and monitored. The physical separation between individual telescope elements adds another level of complexity to these additional calibration requirements. Furthermore, if the telescope elements are spread over a large enough area, additional uncertainties, such as spatially dependent environmental factors, need to be characterised and calibrated \cite{2018Brown}. 
	
	Ground-based $\gamma$-ray telescopes, often referred to as Imaging Atmospheric Cherenkov Telescopes (IACTs), study photons in the energy range of $\sim50$ GeV to above $100$ TeV. At these extreme photon energies, the Earth's atmosphere is opaque. As such, rather than observing the $\gamma$-ray photons directly, IACTs infer their direction and energy from the intensity, temporal and spatial distribution of the Cherenkov radiation emitted by the relativistic leptons in the Extended Air Shower (EAS) created as the atmosphere absorbs the original $\gamma$-ray. For practical reasons, IACTs have no housing for their structures which means that they are continually exposed to the elements and therefore are subject to the effects of weathering. These weathering effects degrade the mirrors and the other optical elements of the IACTs, reducing the reflective efficiency and thus reducing the amount of light transmitted through the IACT optical system. Since the energy of the initial $\gamma$-rays is inferred from the intensity of Cherenkov radiation observed, an unmonitored change in the optical throughput of the telescope system will introduce an uncertainty in the inferred energy of the $\gamma$-rays. Furthermore, if there is a wavelength dependency to the degradation of these mirrors, an additional uncertainty will be introduced since it will have a different impact on the observed signal depending upon which camera technology has been used \cite{2019ApJS..243...11G}.
	
	The Cherenkov Telescope Array (CTA) is the next generation of ground-based $\gamma$-ray telescope array \cite{2013APh....43....3A}. Compared to the current generation of IACT arrays, CTA will provide us with at least an order of magnitude improvement in sensitivity, with unprecedented angular and energy resolution. CTA will also expand the observable energy range of the ground-based technique, with an envisaged energy threshold of 20 GeV and sensitivity beyond 100 TeV. CTA's envisaged improvement in performance will be due to two key factors: (i) CTA will be comprised of three telescope size classes optimised to observe different photon energies ranges and critically (ii) CTA will consist of a total of 120 telescopes spread over five square kilometers. To allow for all-sky coverage, the CTA observatory will consist of two arrays, one in each hemisphere. The northern array is intended to contain $\sim20$ telescopes spread over about one km$^2$, while the southern array is intended to contain $\sim100$ telescopes spread over an area of approximately four square kilometres. 

	While the shear size of CTA will afford us unprecedented accuracy and sensitivity, it will also introduce new calibration challenges; challenges which a UAV-based calibration system has the potential to address. The atmosphere is a critical part of the ground-based $\gamma$-ray detection process; a scientific payload of environmental sensors such as a nephelometer, temperature, humidity and pressure sensors, allows us to quantify the atmosphere above the telescopes, thus allowing us to constrain the uncertainty of atmospheric extinction of the Cherenkov radiation created by the EAS. A UAV-based calibration system also has a unique ability to characterise the optical throughput for all telescopes in the array \cite{2018Brown}. This procedure entails placing a well-understood calibration light source, capable of pulsed illumination on nanosecond timescales, at altitude above the IACT array, and simultaneously illuminating numerous telescopes \cite{2018Brown}. The use of a UAV in this technique, as opposed to a light source on the ground, affords us the possibility of rapidly and simultaneously illuminating all telescopes in CTA.	
	
	The largest IACT array of the current generation is the H.E.S.S. telescope array in the Khomas Highlands of Namibia. Located at an altitude of \SI{1800}{\meter} above sea level, H.E.S.S. consists of five telescopes. Four of them, CT1-4 (HESS-I), have \SI{12}{\meter} dish diameter and are located on a square with side length \SI{120}{\meter}. The larger telescope, CT5 (HESS-II), has \SI{28}{\meter} dish diameter and is located at the centre of this square. In the context of a `proof-of-concept' for the UAV-based airborne cross-calibration of IACT arrays, as outlined in \cite{2018Brown}, H.E.S.S. is a natural test-bed. Of the current generation of IACTs, H.E.S.S. has the largest number of telescopes, and furthermore, it possesses multiple telescope size-classes, analogous to CTA's `Large' and `Medium' sized telescopes. 
	
	In this paper we report the results of the first ever UAV-based, airborne inter-calibration of an IACT array. In Section 2 we outline the test-campaign that was undertaken during May 2018. In Section 3 we present the analysis of the data recorded by the four HESS-I telescopes, including event selection, data cleaning, the position determination of the UAV and the inter-calibration procedure. In Section 4 we outline the telescope response simulations undertaken for the campaign in order to verify the behaviour and optimise the UAV flight profile used during the test campaign. Section 5 presents the first results of the inter-calibration of telescope sensitivity as well as outlining the ability of the UAV-based approach to monitor the pointing accuracy of IACTs. In Section 6, we make our conclusions, and outline the next steps for this new calibration technique. 
	
\section{UAV-Calibration Prototype and Test-Campaign}
\label{TC}
The first test-campaign of our UAV-based calibration system occurred in May 2018, at the H.E.S.S. telescope array. Our first-generation UAV-calibration prototype consisted of a rotary UAV system housing a calibration payload. The rotary UAV was a commercially available, off-the-shelf octocopter powered by a single 16 Ah LiPo battery and positioned with a standard avionics suite consisting of gyroscopes, accelerometers and a Global Navigation Satellite System (GNSS) receiver. The calibration payload was a bespoke LED-based pulsed light source, consisting of four LED boards housing 10 BIVAR UV3TZ-400-15 LEDs each. All four LED boards were placed in a square configuration aligned along the same optical axis, with each board illuminating a $50^{\circ}$ circular top-hat diffuser\footnote{We note that, whilst at a distance of \SI{820}{\meter} from the telescopes, a $20^{\circ}$ would have been sufficient to illuminate all telescopes in the array, a $50^{\circ}$ was used in a conservative approach to negate any adverse affect of inaccurate UAV pointing. While the uniformity of the diffuser was not characterised before deployment, we note that over the $\sim8^{\circ}$ angular extent of the H.E.S.S. array as seen by the UAV, the diffuser intensity is reported to be uniform to within \SI{<5}{\%} \cite{diffuser}. Given that the light source consists of four independent diffusers in a random orientation, we expect a \SI{<2-3}{\%} variation in the intensity of the calibration light as seen by the individual telescopes due to the diffuser's performance. As such, the transmission properties of the calibration light source diffusers are potentially one of the dominant sources of systematic uncertainty of this first campaign, and will be addressed in a later campaign.}. The maximum physical separation between LEDs on the calibration payload was $10$ cm, which can be considered point-like for the UAV-flight profile used in this test-campaign. The individual calibration pulses were triggered by the timing signal from a dedicated GNSS positioning system with a trigger rate of 1~Hz. Characterisation of the light-source before confirmed that the calibration pulses were $4.0\pm0.1$ ns in duration, across a large dynamic range, and that over small ambient temperature variations, the intensity of the calibration light-source did not vary\cite{BrownICRC}. We note however that the goal of this test-campaign was the inter-calibration of the H.E.S.S. array, which does not necessitate a detailed knowledge of the light pulse characteristics.
	
During this first test-campaign, all UAV operations occurred at the H.E.S.S. residence, \SI{800}{\meter} south-east of the telescope array, using a simply vertical flight profile to a maximum altitude of \SI{200}{\meter} from the take-off location, at which point the calibration light pulse fired horizontally. The reasoning for this was three-fold: (i) using a UAV take-off point that was horizontally separated from the telescopes allowed us to achieve the required $\mathcal{O}(0.8-1)$ km UAV-telescope separation without climbing to large altitudes which in turn afforded the UAV more flight time at the calibration location and increased the number of UAV-calibration events recorded per flight; (ii) having a take-off location away from the telescope array greatly reduced the requirement of strict light control by the UAV pilot (which is beneficial for the proof-of-concept nature of the first test-flight campaign); (iii) it allowed us to overcome flight profile restrictions which resulted from using an off-the-shelf commercial UAV. 
\par
For each calibration run, all four HESS-I telescopes were pointed at a predetermined position \SI{200}{\meter} above the UAV take-off point, in convergent pointing mode\footnote{The convergent pointing mode was a dedicated bespoke array configuration created specifically for the UAV calibration in which the H.E.S.S. telescopes point towards a fixed position instead of pointing parallel to a position on the sky. Unfortunately, due to tracking criteria for CT5, it was not possible to include CT5 in the convergent pointing configuration for this campaign. This will be addressed in a later flight test-campaign.} as illustrated in Figure \ref{PosDet}. For this first campaign, the telescopes started their observing run as the UAV flew towards the calibration position. This approach maximised the number of calibration events recorded, but also led to a small number of calibration events being recorded at the edge of the cameras as the UAV flew into the field-of-view, and as it exited during the landing phase of the UAV flight profile. To ensure that the calibration light source was aligned with the telescopes, the UAV light source was aligned in the direction of the telescopes before the flight, with the alignment confirmed again at the end of the calibration flight. Furthermore, no pilot yaw input was given to the UAV during the flight, and the UAV's yaw stability in free-flight was confirmed before the field campaign to be $\pm2.5^{\circ}$ \cite{BrownSPIE}. 
\par 
For the UAV-calibration run, standard observation trigger settings were used, with the exception that no constraint was put on the number of triggered telescopes needed to create an event and as such, events triggering only one telescope were accepted. Being at a horizontal separation of about \SI{800}{\meter} from the H.E.S.S. array centre at an altitude of \SI{200}{\meter} led to a separation of \SI{\sim820}{m} between the H.E.S.S. array and the UAV during calibration runs. However, IACTs, such as the H.E.S.S. telescopes, are focused to observe the EAS at shower maximum, \SI{8}{\kilo\meter} above the array \cite{cornils2003}. As expected, the proximity of the UAV relative to the distance at which the IACTs are focused resulted in a smearing of the image of the UAV-based calibration light source due to aberration effects. The magnitude of these aberration effects is dependent upon the depth of field of the individual telescopes and the position of the UAV relative to the telescope’s optical axis. At its worst, the point-like image of the calibration light source will be smeared across multiple pixels, thus increasing the statistical fluctuations due to the contribution of night sky background light to the recorded signal \cite{2018Brown}. It should be noted that with a separation of \SI{\sim820}{m} between the array and the UAV, the calibration images recorded by the HESS-I cameras were smeared beyond one pixel. 
 \par
 In total, two successful UAV-calibration runs were taken during this campaign. The first successful run, henceforth referred to as run A, occurred on the $20^{\text{th}}$ of May 2018 at 21:29 UTC, and recorded a total of nearly \SI{154000}{} events, where the term `events' encompasses UAV-calibration events and cosmic ray induced EAS events (see Figure \ref{EvSelDisp}). The second successful run, henceforth referred to as run B, occurred on the $21^{\text{st}}$ of May 2018 at 22:33 UTC, and recorded a total of nearly \SI{102000}{} events. As will be discussed in Section \ref{EvSel}, $343$ and $350$ of these events are selected as UAV events. Given that run A and run B occurred on different days, the environmental conditions during each successful run were noted and can be seen in Table \ref{environmental}.

\begin{table}
\centering
\begin{tabular}{|c|c|c|}
\hline
 Environmental Parameter & Run A & Run B \\
\hline \hline
 Mean Wind Speed (m/s) & 1.2 & 1.6 \\
 Mean Wind Direction ($^{\circ}$ from North) & 79.3 & 90.4\\ 
 Mean Temperature ($^{\circ}\text{C}$) & $7.07 \pm 0.21$ & $10.61 \pm 0.38$\\
 Mean Humidity ($\%$) & $79.40 \pm 0.45$ & $10.61 \pm 1.79$ \\
 Mean Pressure (mbar) & $823.00 \pm 0.08$ & $825.00 \pm 0.02$\\
 \hline
 \end{tabular}
\caption{Environmental conditions during the successful UAV-calibration runs. The wind speed, temperature, humidity and pressure was recorded by the H.E.S.S. array's weather monitoring system on the ground.}
\label{environmental}
\end{table}


\section{Reconstruction and Analysis of UAV-Calibration Events}
\label{analysis}
Before identifying the UAV-calibration events and extracting the calibration signal, standard data cleaning procedures were applied to all data recorded by the HESS-I cameras. First, we removed the pedestal, which is the electronic baseline, from the charge accumulated in each pixel. The pedestal was determined from the mean night sky background obtained from non-illuminated pixels in each event. After pedestal subtraction, standard gain calibration was applied to the data to convert the charge accumulated in each pixel, to the number of photo-electrons measured by each pixel \cite{hess-calibration}. Thereafter, the expected signal in non-operational pixels\footnote{Non-operational pixels are due to broken hardware preventing a pixel from recording a correct signal or to bright stars passing through the field of view of a pixel leading to the necessity of disabling it as described in \cite{hess-calibration}.} was interpolated from the mean intensity of the neighbouring operational pixels (average over six pixels if all the neighbouring pixels are operational and the pixel is not at the edge of the camera). Finally, the data were cleaned in a `tailcut' cleaning procedure keeping only pixels which fulfilled a dual-threshold condition: the pixel must have recorded at least seven photo-electrons and have a neighbouring pixel with at least five photo-electrons or, the pixel has recorded at least five photo-electrons and a neighbouring pixel at least seven photo-electrons (see \cite{hess-crab} for more details). 
		
\subsection{Event Selection}
	
	\label{EvSel}
	\begin{figure}
		\centering
		\begin{subfigure}{.32\textwidth}
			\centering\includegraphics[width=1.\linewidth]{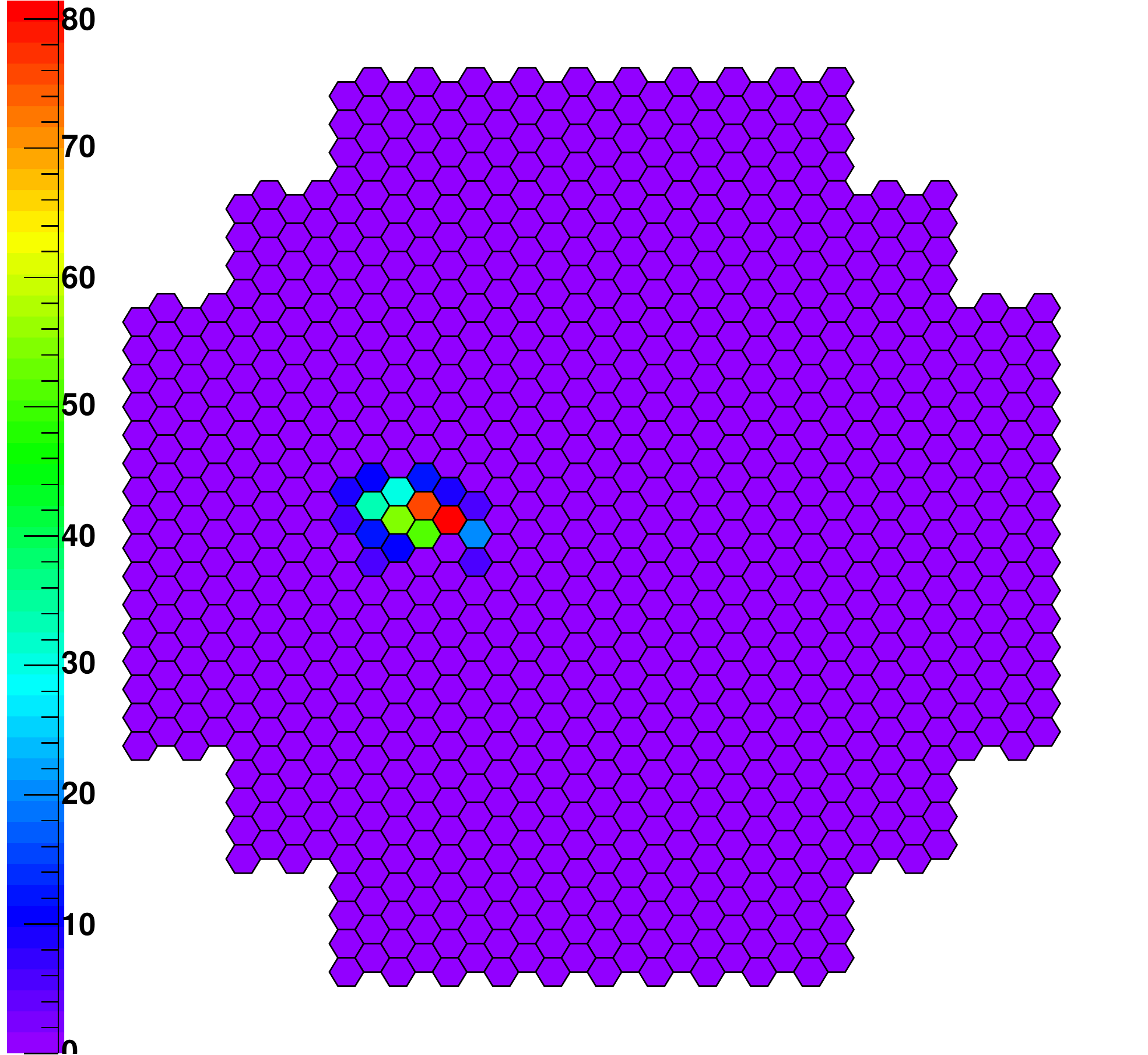}
		\end{subfigure} 
		\begin{subfigure}{.32\textwidth}
			\centering\includegraphics[width=1.\linewidth]{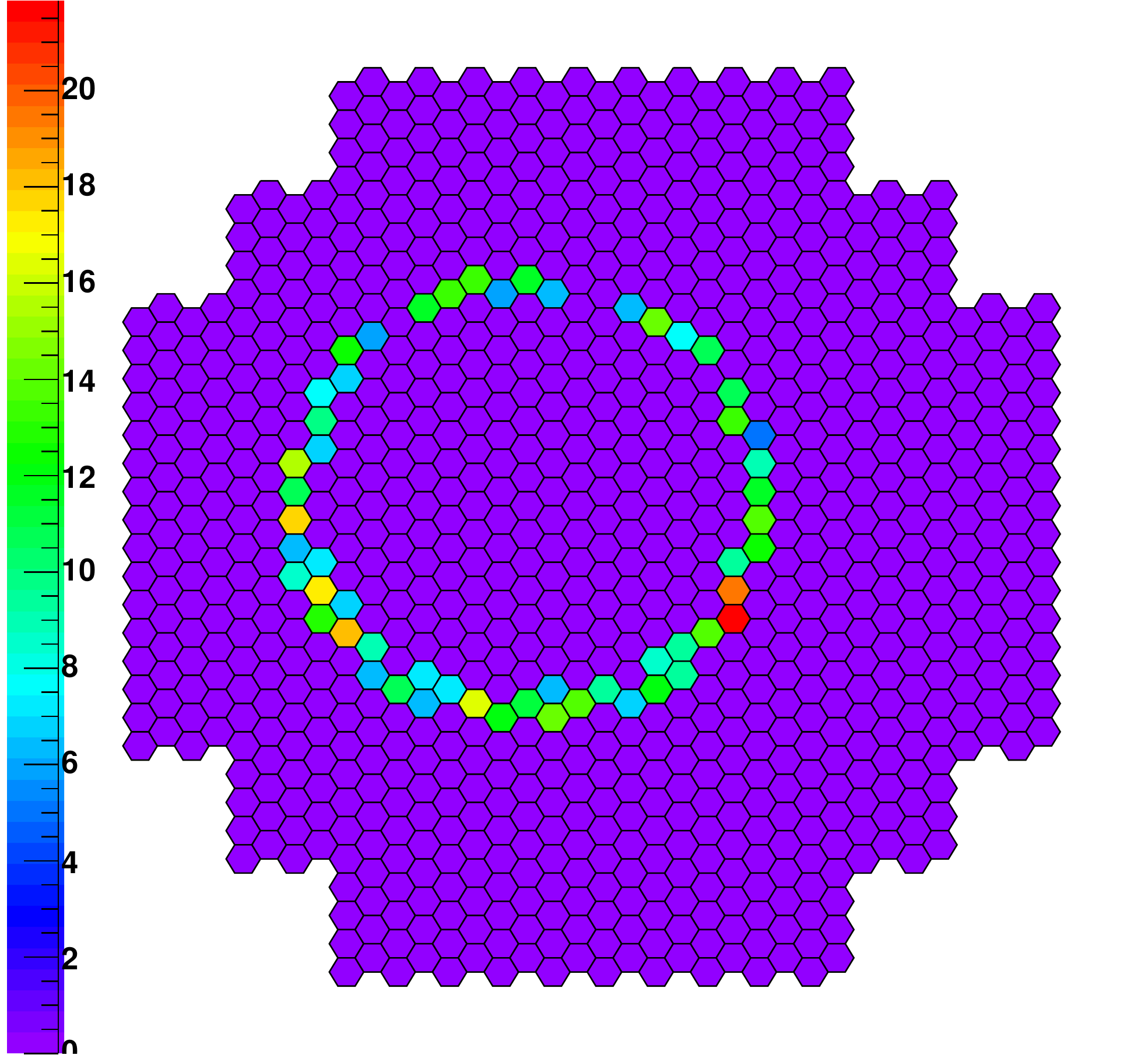}
		\end{subfigure}
		\begin{subfigure}{.32\textwidth}
			\centering\includegraphics[width=1.\linewidth]{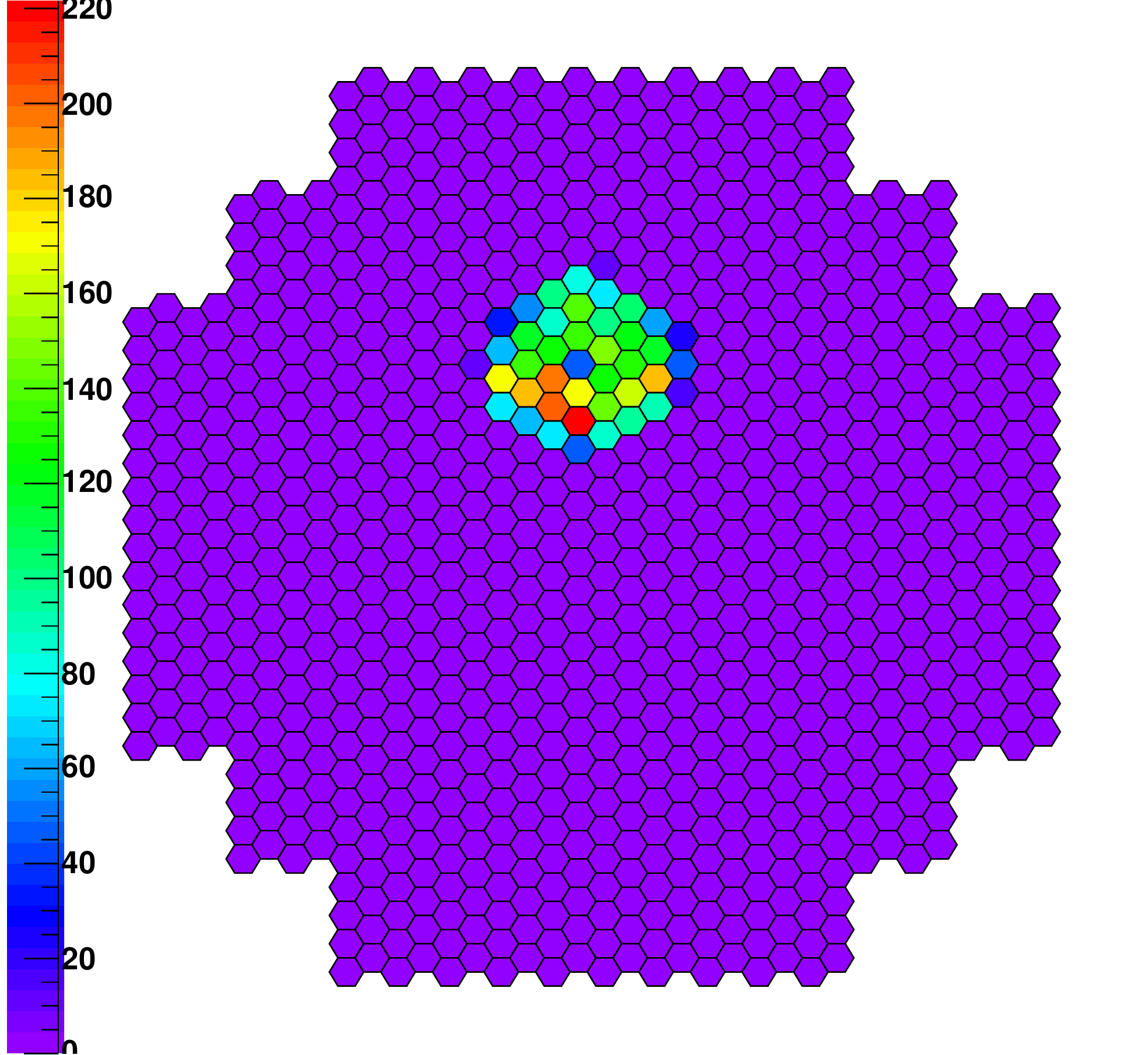}
		\end{subfigure} 
		\caption{Example event displays for the different types of measured events. Left: Cosmic event; Middle: Muon ring; Right: UAV-calibration event. The colour scale indicates the number of photo-electrons recorded in each pixel, with all pixels having a $0.16^{\circ}$ diameter field of view.}
		\label{EvSelDisp}
	\end{figure}
	
Once all events recorded during the UAV-calibration run were cleaned, the UAV-calibration events were identified and retained, with background events being rejected. As can be seen in Figure \ref{EvSelDisp}, three types of events were in the recorded data:
	\begin{itemize}
		\item{Cosmic events: events from high energy photons, electrons or hadrons entering the atmosphere, which are characterised by an elliptical or irregular shape. These cosmic events were recorded by one or two telescopes due to the telescopes operating with convergent pointing at low altitude.}
		\item{Muon rings: Cherenkov rings produced by atmospheric muons (also originating from cosmic events) crossing the telescope characterised by their ring form. Muon rings are single telescope events.}
		\item{UAV events: characterised by their regular hexagonal shape and the fact that they are predominatedly recorded by four telescopes. Additionally, UAV events are much brighter than most cosmic events.}
	\end{itemize}
	\begin{figure}
	    \centering\includegraphics[width=11cm]{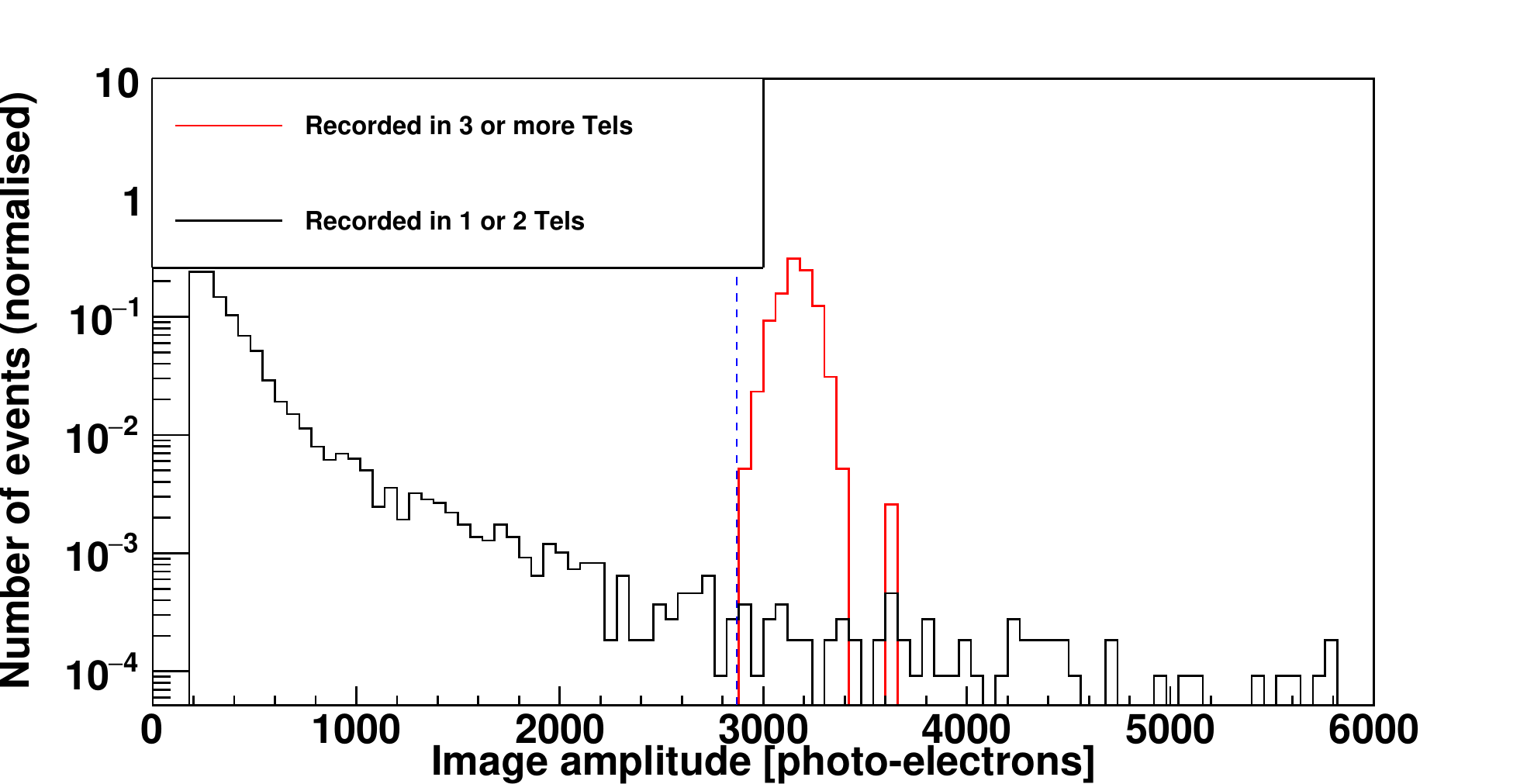}
		\caption{Normalised distribution of the image amplitudes for events recorded in one or two telescopes (black) and for events recorded in three or four telescopes (red) on logarithmic scale as example for CT2 and run A. It can be seen that no event being recorded by 3 or 4 telescopes and so classified as UAV events has an image image amplitude below \SI{2870}{} photo-electrons (illustrated by the blue dashed line). However, the distribution of background cosmic events extends beyond this image amplitude threshold.}
		\label{SelCuts}
	\end{figure}

	After removing events which were not completely contained on the camera focal plane, by applying a cut of \SI{0.034}{\radian} (\SI{\approx2}{\degree}) on the angular distance between the centre of the camera and the image centre of gravity, events having been recorded in at least three telescopes were selected as UAV events. 
	\par 
	To verify the performance of this cut, the UAV was set up such that UAV events also have a higher image amplitude than most cosmic events as illustrated in Figure \ref{SelCuts}. In total, there were six runs taken under very similar conditions (similar pointing directions, weather conditions, ...) in the UAV campaign, whereof two were successful i.e. there were UAV events completely recorded by the cameras of all four HESS-I telescopes. Of the four remaining runs, one had a user-error resulting in a calibration pulse duration that was too long compared to the integration time of the HESS-I cameras, in two runs the UAV did not completely enter the field of view of the telescopes due to an inaccurate coordinate transformation and the last unsuccessful run only had three telescopes recording data. In the three unsuccessful 4-telescope runs not considered in the main analysis, there were in total \SI{242324}{} events being classified as cosmic events in all telescopes in which they were recorded. There were, however, no events recorded in three or four telescopes and not rejected based on nominal distance in these runs, except for the run in which the UAV had a too long pulse duration. The longer pulse duration however also led to a higher image amplitude and so all the events recorded in three or four telescopes in this run had image amplitudes above \SI{20000}{} photo-electrons, except for two events which turned out to be clearly caused by the UAV on visual inspection. Making use of the fact that there are \SI{242324}{} events clearly identified as cosmic events (i.e. events having triggered one or two telescopes and having an image amplitude of less than \SI{20000}{} photo-electrons for the run with the long pulse duration) in these three runs and that there are \SI{148367}{} and \SI{97976}{} events recorded in one or two telescopes in the good runs A and B respectively and assuming that the probability $p$ for a cosmic event to be recorded in three or four telescopes follows a binomial distribution, there are no cosmic events recorded in three or four telescopes with a confidence of \SI{80.4}{\%} respectively \SI{91.5}{\%} in both of these runs (corresponding to p-values of $0.196$ and of $0.085$ respectively). This shows that it is very unlikely to have any cosmic event misclassified as UAV event using the cut based on telescope multiplicity and for this reason this is the cut retained for the UAV event selection.
	\par 
	To further confirm the nature of the calibration events identified as UAV events, we considered the time stamps of all events which pass the UAV selection criteria. The time stamps of all UAV-selected events were found to be a multiple of the GNSS pulse period used to trigger the calibration events, increasing the confidence that no cosmic event has been wrongly selected as UAV events. Furthermore, from this time stamp information, we note that the small number of UAV events not recorded by the array trigger, despite the UAV being in the field of view of all four telescopes, was consistent with the expected dead-time of the H.E.S.S. telescope system or part of it. Finally we note that the timing was not used for the event selection in order to have an event selection independent of the timing pattern of the UAV in order to be able to easily adapt this pattern and as using the timing is not expected to lead to a more efficient and accurate event selection.
	\par 
	\begin{figure}
		\centering
		\begin{subfigure}{.9\textwidth}
			\centering\includegraphics[width=1.\linewidth]{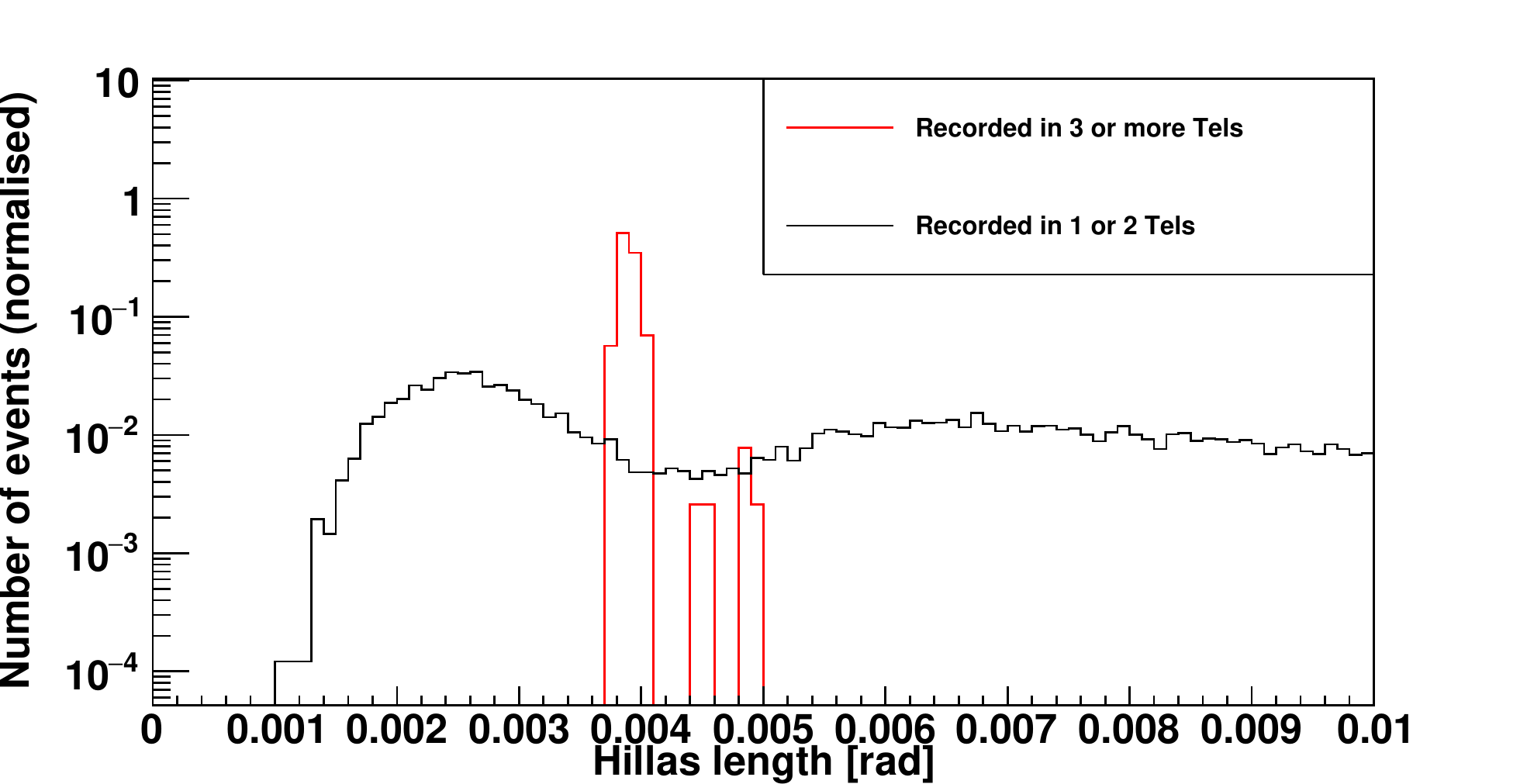}
		\end{subfigure} 
		\begin{subfigure}{.9\textwidth}
			\centering\includegraphics[width=1.\linewidth]{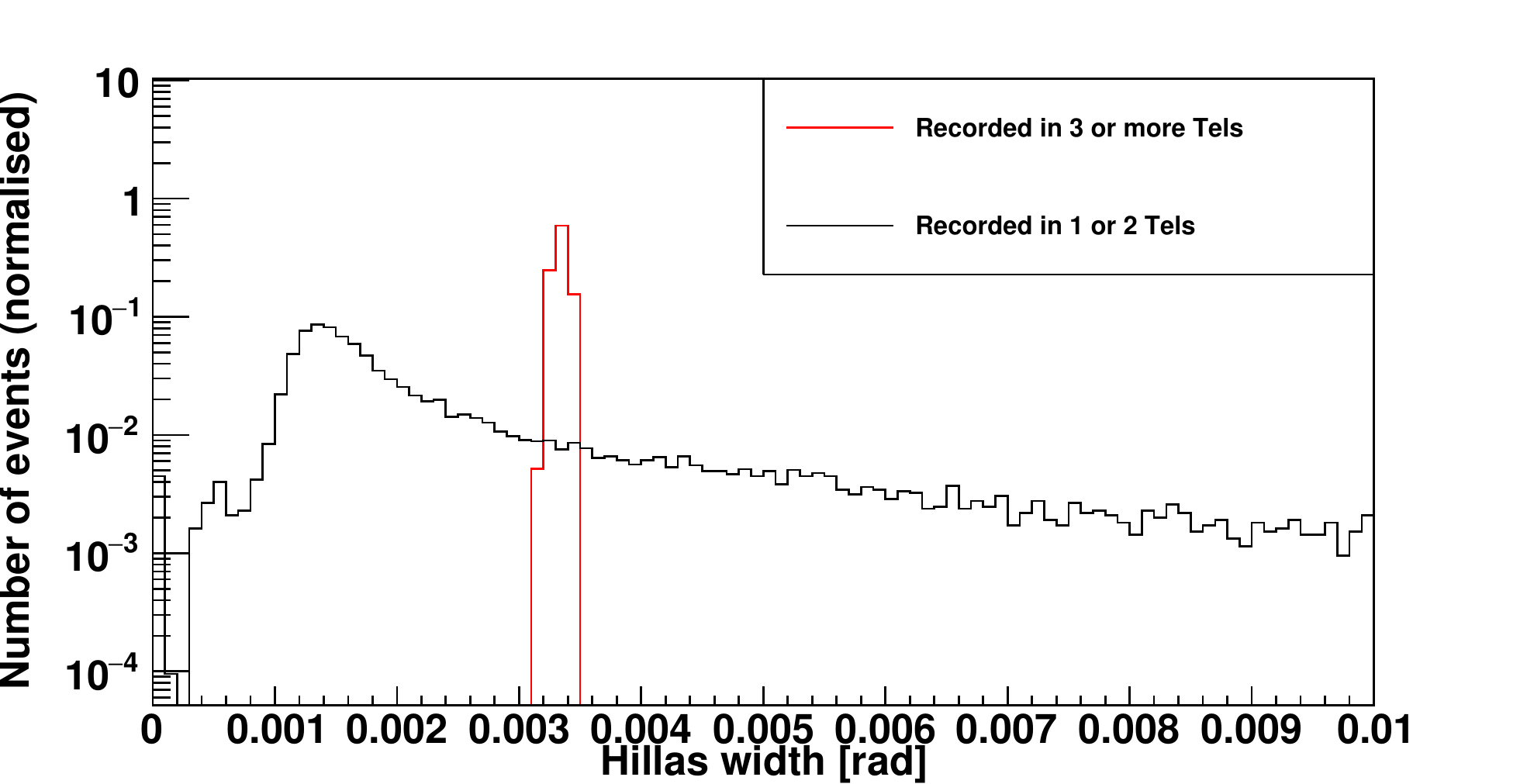}
		\end{subfigure}
		\begin{subfigure}{.9\textwidth}
			\centering\includegraphics[width=1.\linewidth]{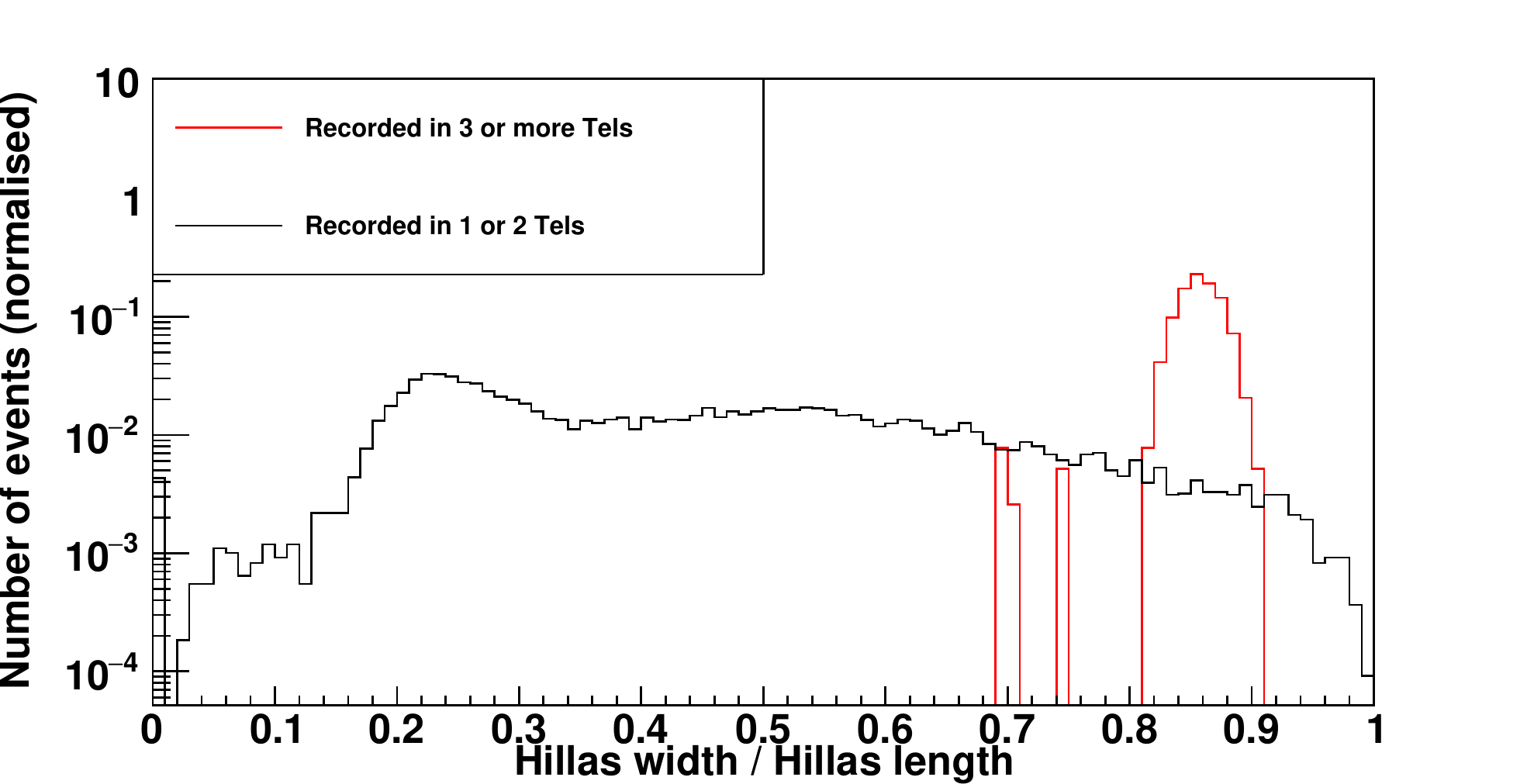}
		\end{subfigure}
		\caption{Normalised distribution of other possible selection variables for events recorded in one or two telescopes (black) and for events recorded in three or four telescopes (red) on logarithmic scale as example for CT2 and run A. Top: Hillas length; Centre: Hillas width; Bottom: Hillas width divided by the Hillas length.}
		\label{OtherPosCutVar}
	\end{figure}
	For completeness, other possible discrimination variables, based on the so-called Hillas parameters which are traditionally used to parametrise images of gamma-ray showers in Cherenkov telescopes \cite{Hillas}, were considered. In this parametrisation, the image of a shower is modelled by a two-dimensional ellipse with a given major axis (called Hillas length) and minor axis (Hillas width). The distributions of these both variables as well as the Hillas width divided by the Hillas length for events which are recorded in one or two telescopes and for events recorded in three or four telescopes are shown in Figure \ref{OtherPosCutVar} as example for run A. Even though all three variables tend to take lower values for events triggering one or two telescopes, the distributions are still overlapping and so not so well suited for discrimination as the number of telescopes in which an event is recorded.
	\par
	For run A, \SI{343}{} events out of a total of \SI{154000}{} were selected as UAV-calibration events, while from run B, a total of 350 out of \SI{102000}{} were selected as UAV events.
	\subsection{Determination of UAV Position}
	\label{PosDetSection}
	\begin{figure}
		\centering
		\begin{subfigure}{1.\textwidth}
			\centering\includegraphics[width=.8\linewidth]{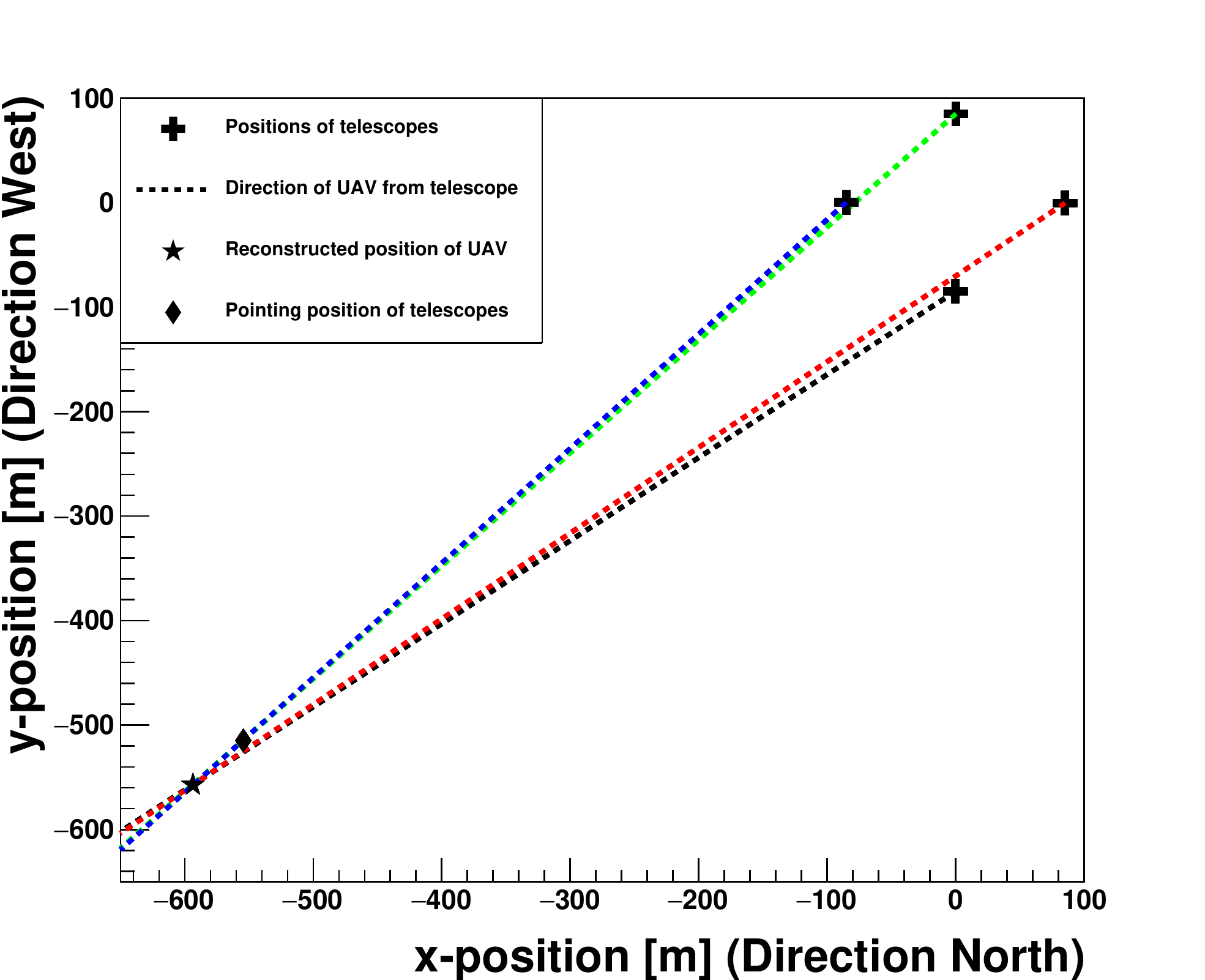}
	\end{subfigure} 
		\begin{subfigure}{1.\textwidth}
			\centering\includegraphics[width=.8\linewidth]{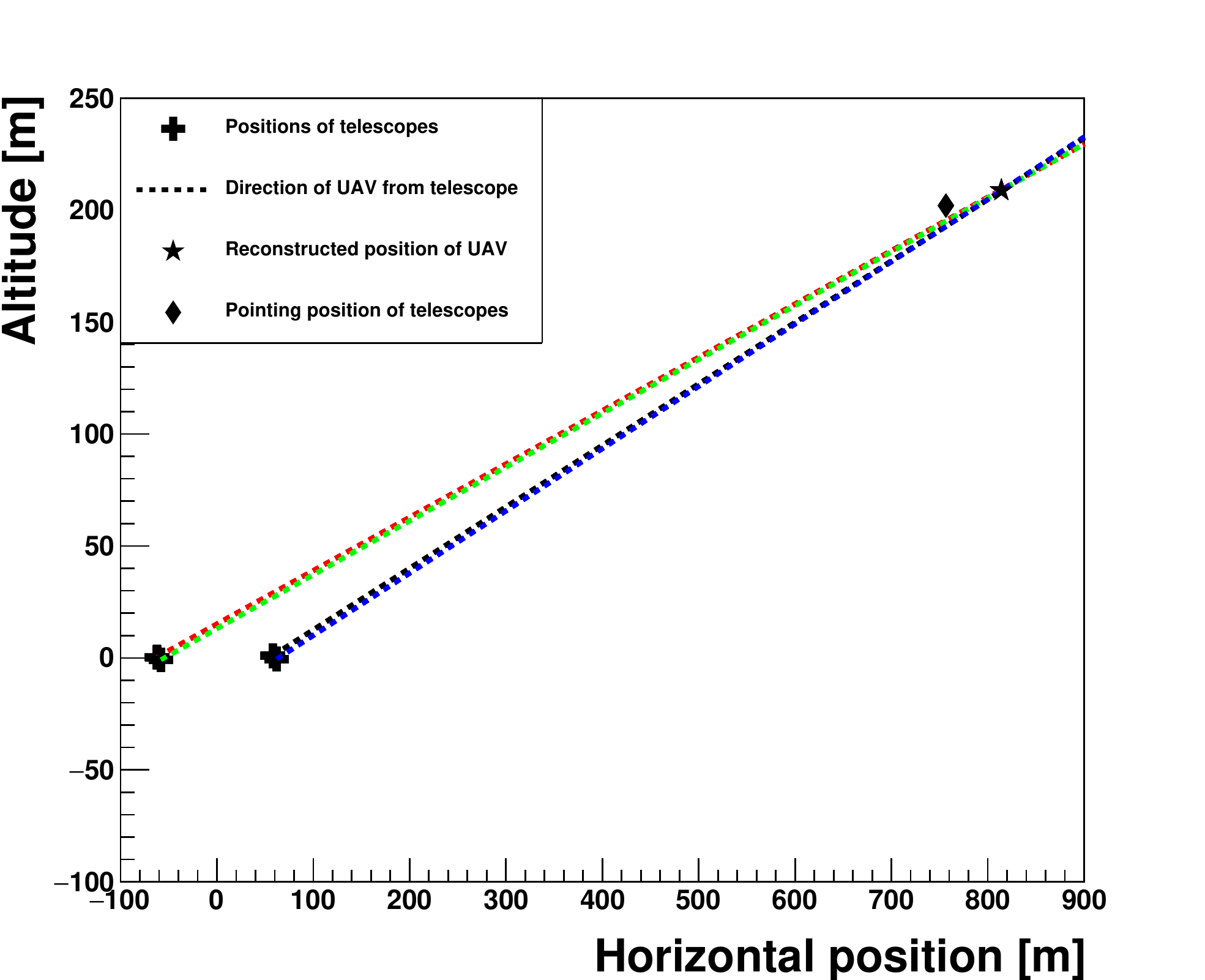}
		\end{subfigure} 
		\caption{Projection on the ground plane (bird-eye perspective) (top) and the vertical plane defined by the UAV and the centre of the array (ground-observer perspective) (bottom) of the situation to illustrate the position determination of the UAV. The crosses indicate the position of the telescopes, the lines the direction in which the UAV has to be with respect to each telescope and the star the point closest to an intersection of the lines i.e. the reconstructed UAV position. The diamond indicates the pointing position of the telescopes.}
		\label{PosDet}
	\end{figure}
	\label{UAVPos}
	Due to drift in the satellite navigation, pilot input and buffeting from atmospheric turbulence, the UAV was moving in the field of view during each calibration run. This makes it necessary to have a precise tracking system of the UAV to determine its position and thus its distance to the different telescopes, which is a crucial value for the inter-calibration technique. The position of the UAV was computed by triangulation using the images of the light source on the camera focal plane: specifically by convolving the centre of gravity of the image with the known pointing direction of the telescope we can deduce the altitude and azimuthal angle of the UAV in the reference frame of the telescope. This leads to a direction in which the UAV is for each telescope having recorded the event (situation illustrated in Figure \ref{PosDet}). If the directions were perfectly determined, the lines defined by these directions would intersect in the point corresponding to the UAV location. However, among other due to statistical variations in the images, the lines do not intersect and so the most likely position of the UAV is taken to be the analytically determined point in space with the minimum sum of squared distances to the lines of sight. With this method, we found that the UAV moved up to \SI{30}{\meter} in altitude (mainly while entering and leaving the field of view due to pilot input to optimise the image position on the camera focal plane\footnote{It is worth to note that no cut on the UAV position beside the rejections of events which were not fully included in the camera based on the nominal distance cut is used. For this reason, also the events recorded while the UAV is entering and leaving the field of view are considered.}) and less than \SI{5}{\meter} in horizontal direction. 
	\par 
	\begin{figure}
		\centering
		\begin{subfigure}{.49\textwidth}
			\centering\includegraphics[width=1.\linewidth]{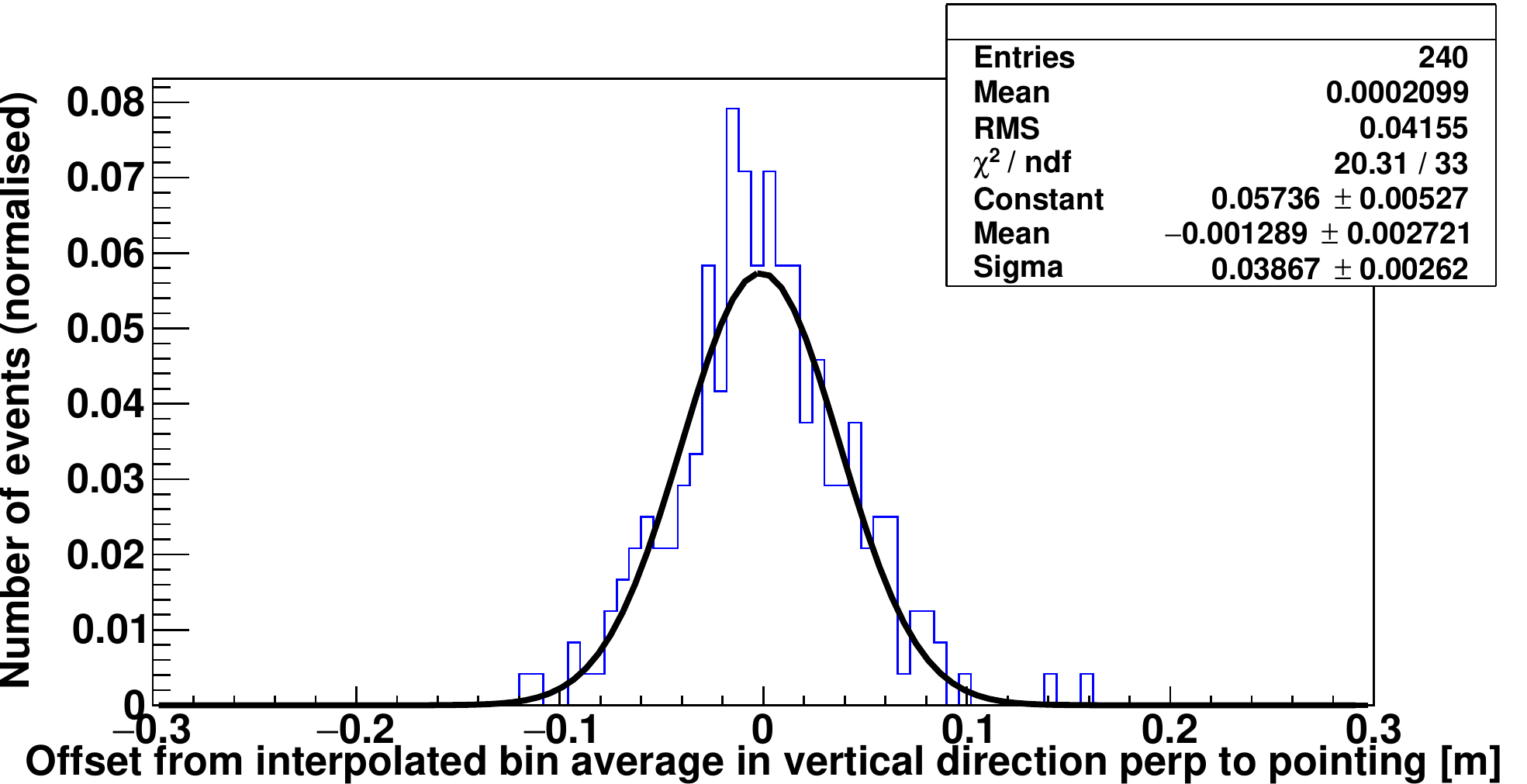}
		\end{subfigure} 
		\begin{subfigure}{.49\textwidth}
			\centering\includegraphics[width=1.\linewidth]{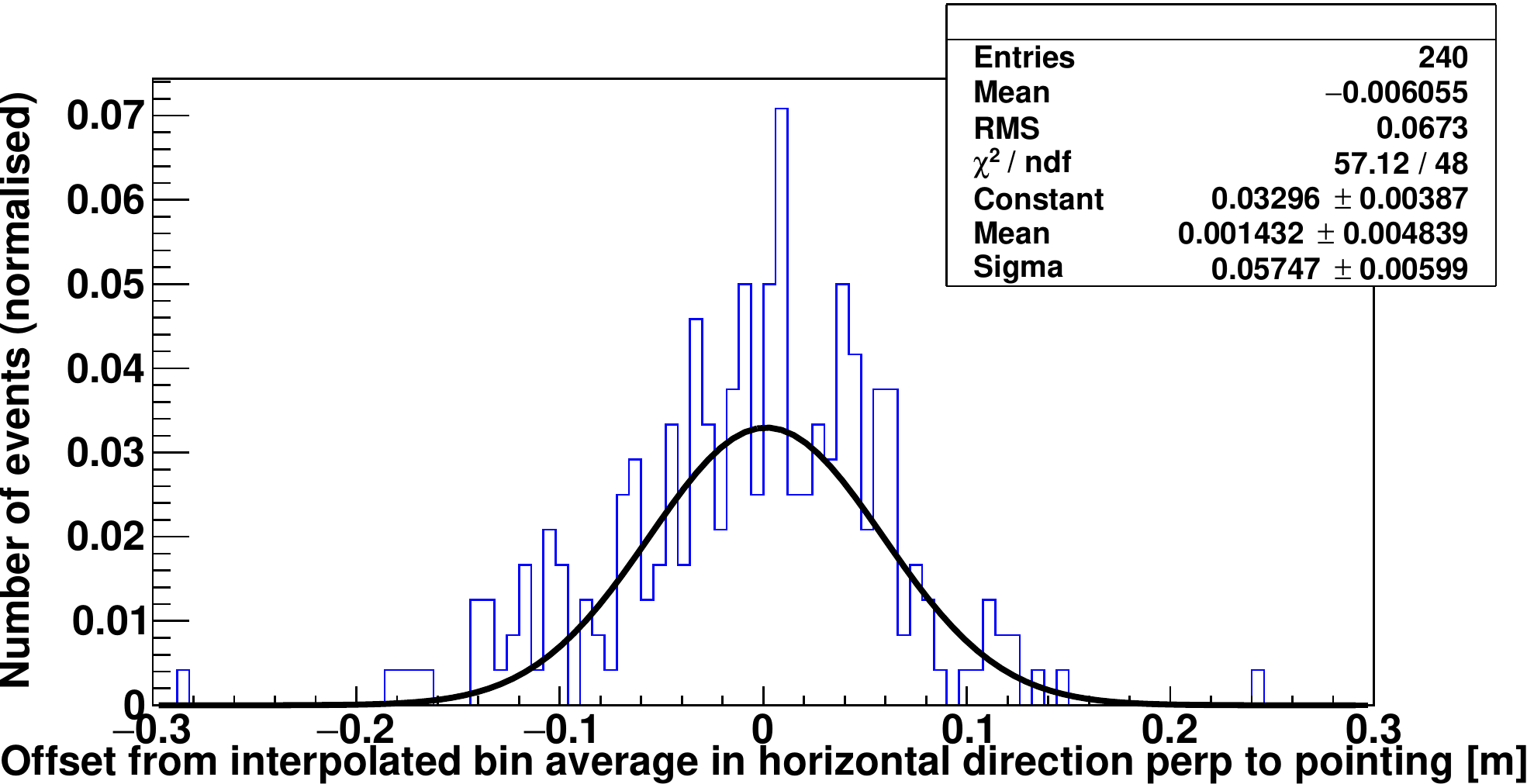}
		\end{subfigure}
		\begin{subfigure}{.49\textwidth}
			\centering\includegraphics[width=1.\linewidth]{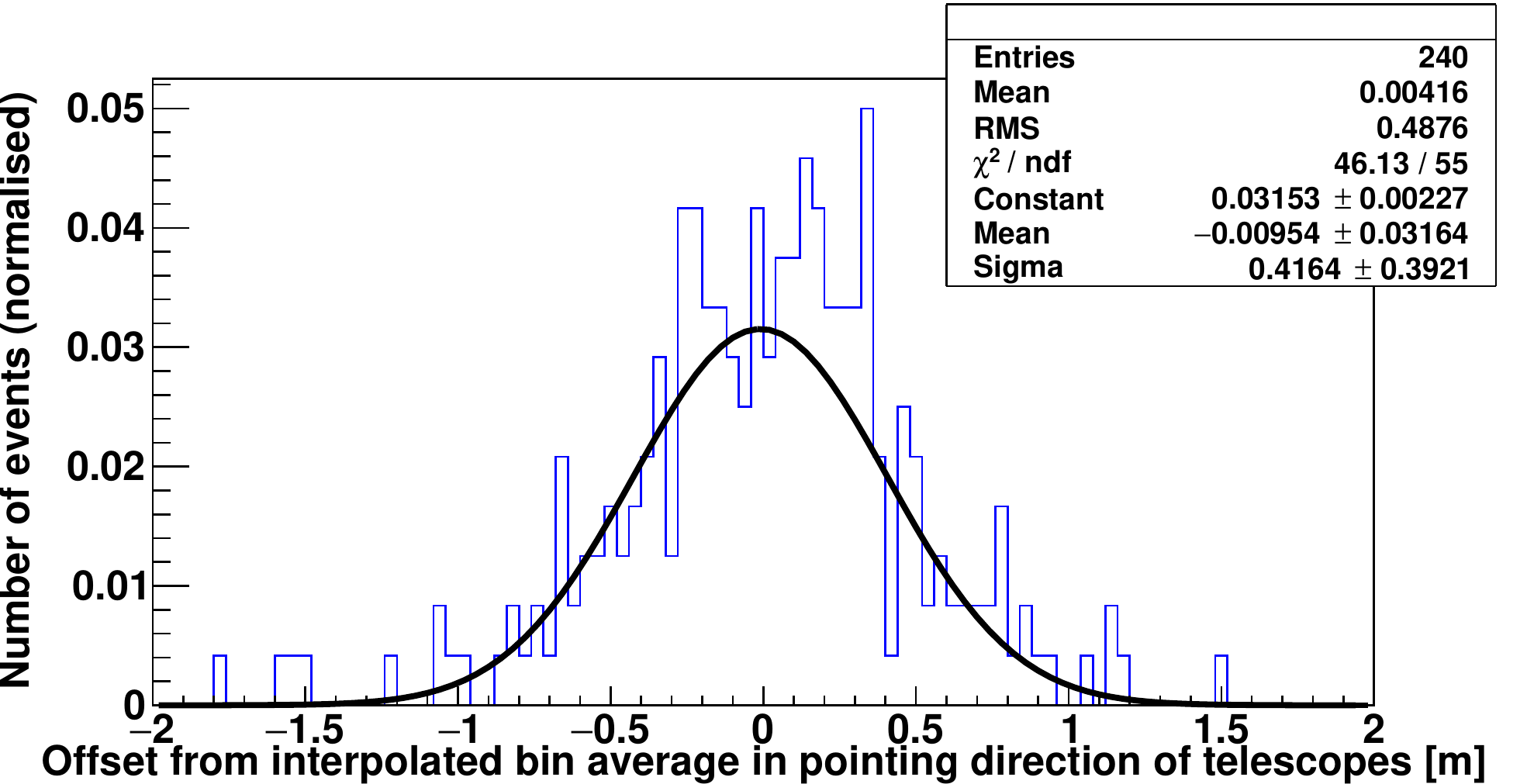}
		\end{subfigure}
		\caption{Distribution of offset in reconstructed UAV position from linearly interpolated 5-second bin average reconstructed position for run A. As only bins with $5$ UAV events and whose neighbouring bins have $5$ UAV events were used, these distributions contain only $240$ events of the about $400$ UAV events recorded in at least three telescopes in run A. Top left: In vertical direction perpendicular to the pointing direction of the telescopes. Top right: In horizontal direction perpendicular to the pointing direction of the telescopes. Bottom: In pointing direction of telescopes. Note the different scale on the position axis for the bottom plot compared to the two top plots.}
		\label{PosOffset}
	\end{figure}
	To determine the statistical uncertainty of the reconstructed positions, the total duration during which the UAV was visible to at least three telescopes (\SI{400}{\second} to \SI{500}{\second}) was subdivided in bins of \SI{5}{\second} and the average of each position coordinate was computed for each bin containing the expected five events (as the UAV was pulsed with a frequency of \SI{1}{\hertz}). Then, the position coordinates were linearly interpolated between the time bin centres to get an expected UAV position at each time (for the time bins having two filled neighbouring bins, the others are not taken into account here anymore). This then allows to compute the offset of the measured UAV position from the expected `5-second average' position and so to get a handle on the statistical position uncertainty. 
	The distribution of the offset of the reconstructed position from the interpolated bin average reconstructed position is shown on an event-by-event basis in Figure \ref{PosOffset} for run A. It can be seen that the distributions on this Figure \ref{PosOffset} can be approximated with a Gaussian distribution, and the standard deviation of this Gaussian corresponds to the 1-sigma statistical uncertainty of the given position coordinate. This leads to a statistical uncertainty on the determined position in pointing direction of the telescopes of about \SI{40}{\centi\meter} and in directions perpendicular to the pointing direction of about \SI{5}{\centi\meter} (per axis) being equivalent to an angular uncertainty of \SI{12.3}{\arcsecond}. That the uncertainty is much lower in the direction perpendicular to pointing was to be expected: in fact, all the telescopes are pointing almost in the same direction and so the lines (defined by the direction in which the UAV is for each telescope) are close to each other in this direction over a long distance (see Figure \ref{PosDet} for illustration) which leads to a less precise determination of the coordinate in this direction than in the directions perpendicular to it. For run B, the UAV left and reentered the field of view thrice during data taking leading to quick movements and accelerations, even on scales as small as \SI{5}{\second}. A more or less stable UAV position was only reached before the first exit of the UAV out of the field of view. Repeating the same procedure as before on this short time interval of \SI{35}{\second}, the same uncertainties of \SI{5}{\centi\meter} and \SI{40}{\centi\meter} respectively were found, this time based on $35$ events, and so with a much lower statistics than the previous $240$ events.
	\par
	\begin{figure}
		\centering
		\begin{subfigure}{.49\textwidth}
			\centering\includegraphics[width=1.\linewidth]{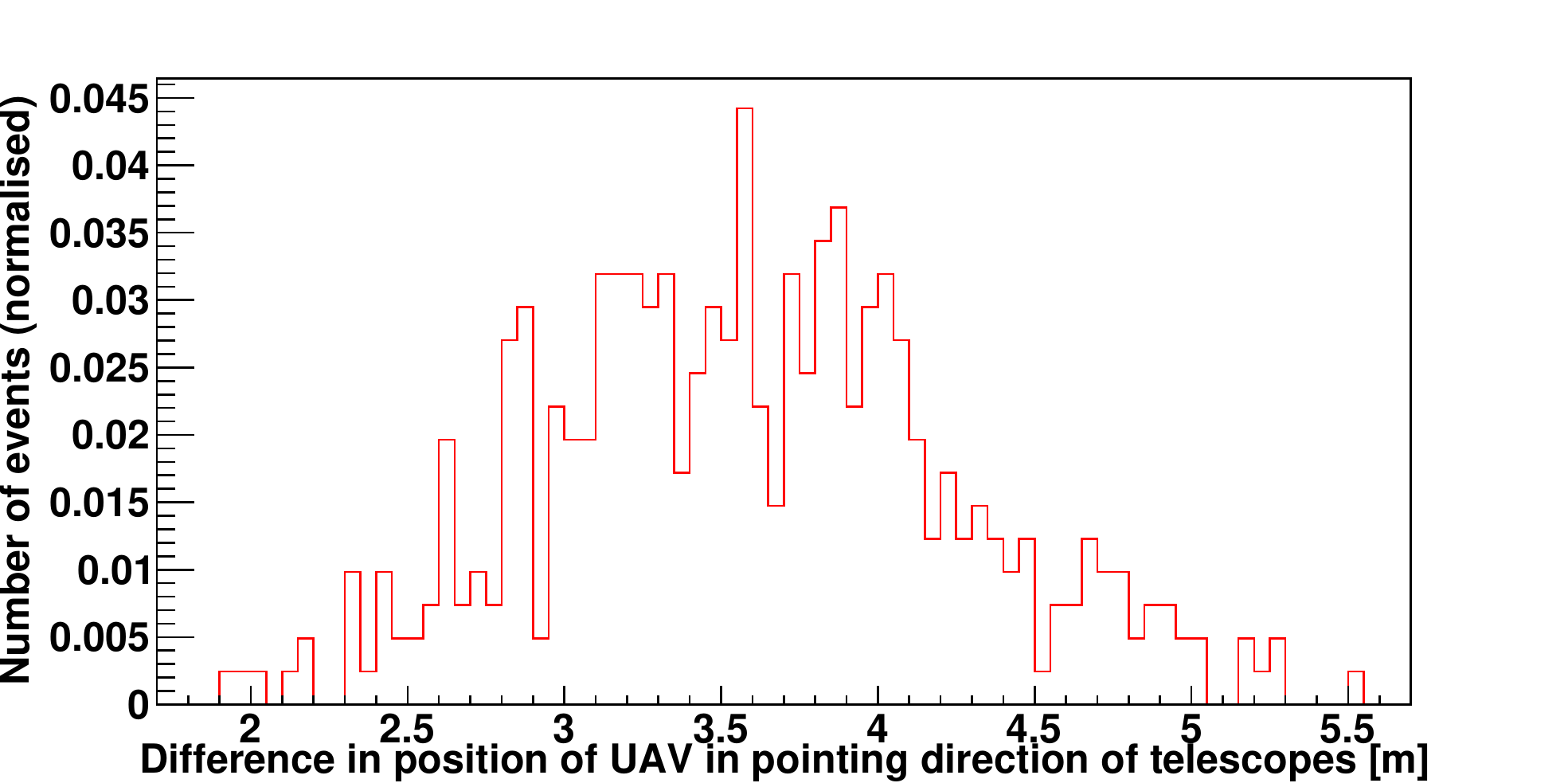}
		\end{subfigure} 
		\begin{subfigure}{.49\textwidth}
			\centering\includegraphics[width=1.\linewidth]{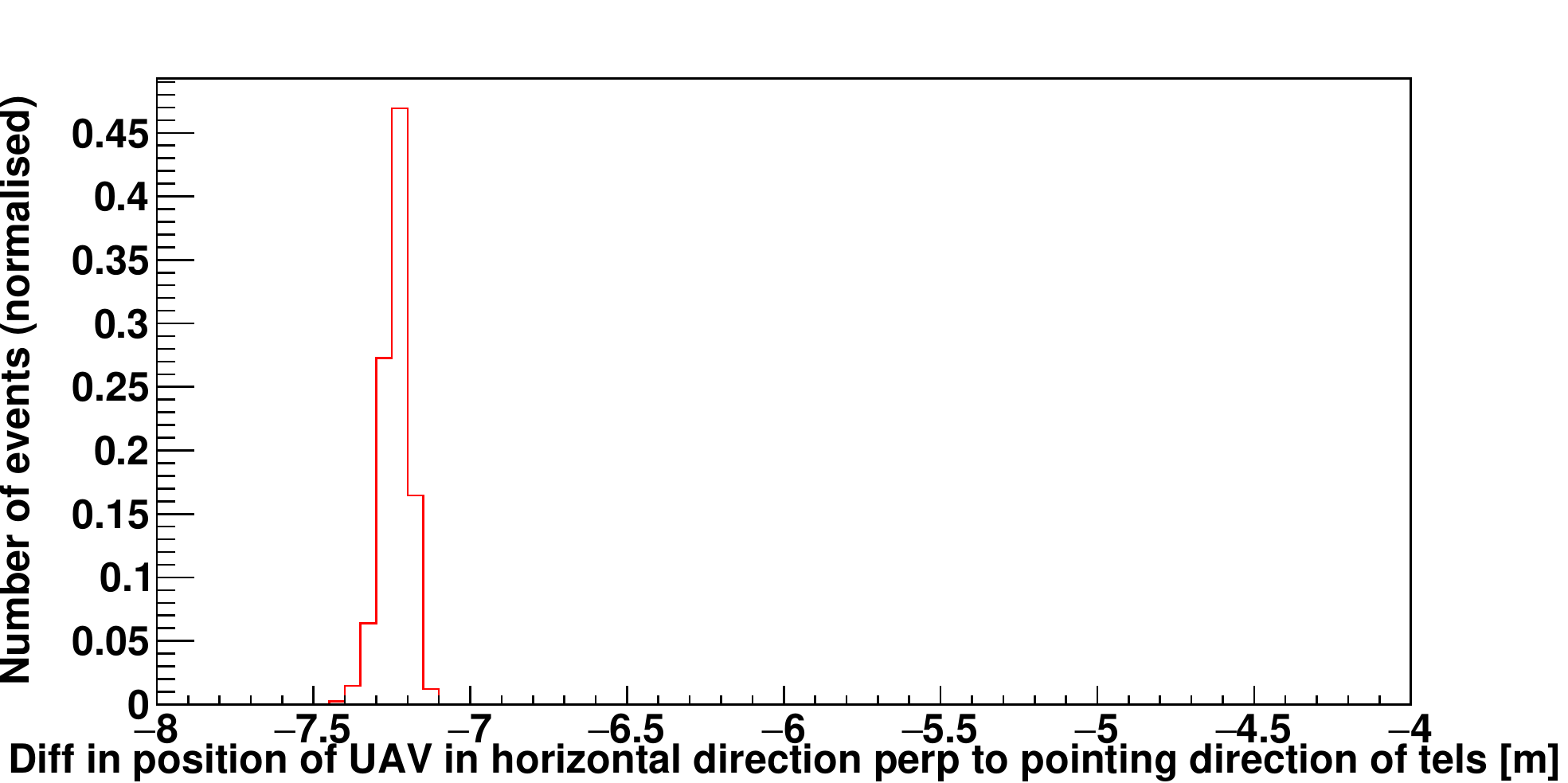}
		\end{subfigure}
		\caption{Distribution of the differences of the determined position coordinates with the field of view method and the GNSS on an event-by-event basis for run B. Left: In average pointing direction of the telescopes (As the altitude of the H.E.S.S. telescope array is not known with precision, any possible offset of the altitude is not included in this plot. The altitude of the H.E.S.S. array was set such that it is consistent with the average altitude found with the FoV method.); Right: In direction perpendicular to average pointing direction in horizontal plane. The axis ranges have been chosen in such a way that both cover the same length in order to emphasise that the distribution on the right is much narrower than the distribution on the left.}
		\label{FOVGPSDiff}
	\end{figure}
	To get the order of magnitude of possible systematic uncertainties of this Field of View (FoV) method, its results were compared to a second completely independent method with no common systematic uncertainties: The second position determination was based on the GNSS mounted on the UAV to enable it to follow a predefined track. Figure \ref{FOVGPSDiff} shows the difference between the positions obtained with the FoV method and the GNSS on an event-by-event basis for run B. Run A is missing, because the GNSS data could unfortunately not be recovered for this run. The maximum difference is \SI{8}{\meter} or less for each coordinate. This difference is composed of a constant shift and a component varying with the reconstructed $d$. Perpendicular to pointing direction, the shift is \SI{7}{\meter} to \SI{8}{\meter} and the spread is very low (about \SI{7}{\centi\meter}). In the pointing direction, the mean is also shifted by about \SI{3.5}{\meter}, but moreover the spread is now much higher reaching \SI{1.1}{\meter}. The reason for this shift is not clear. It could come from systematics in the GNSS method (which are expected to be up to \SI{10}{\meter}), inaccurate knowledge of the position of the centre of the H.E.S.S. array in the GNSS system or systematics in the field of view method. This means that the difference could be dominated by the uncertainty in the GNSS position determination and that the uncertainty on the FoV method is much lower than \SI{8}{\meter}. However, it is not possible to constrain the uncertainty on the FoV method further than the uncertainty on the GNSS method by comparing it to this method. This shows that the systematic uncertainty on the FoV method is at maximum of the order of magnitude of \SI{8}{\meter} (per axis), but could be much lower down to \SI{40}{\centi\meter} in pointing direction and \SI{5}{\centi\meter} perpendicular to pointing direction. These uncertainties (as well as the statistical ones) hold for this particular geometry (modulo the absence of any absolute altitude comparison) and might be different for other geometries, even though the used geometry has no particularity from which particular low or high uncertainties would be expected.

	\subsection{Inter-Calibration of the Telescopes}
	\label{inter-calibration}
	To inter-calibrate the HESS-I telescopes, only UAV events which were recorded in all four telescopes were considered, and for each event $I\times d^2\times C$ was computed, where $I$ is the sum of all the photo-electrons in an interpolated and cleaned event in a given telescope, $d$ the distance of the UAV to the mirror plane (i.e. the plane perpendicular to the telescope pointing direction containing the centre of the mirror) of this telescope and $C$ a correction factor close to $1$ accounting for atmospheric extinction and higher order geometric effects. As will be shown by our UAV-specific simulations in Section \ref{Sim}, for the distances that the UAV was from CT1-4 during the calibration runs, $I\times d^2$ is the same for all telescopes, modulo small percentage or lower level variations due to atmospheric extinction, point-to-point variations and higher order geometric effects. The atmospheric extinction and the next order geometric effect accounting for UAV movements perpendicular to the mirror axis and the finite mirror size have been implemented in the correction factor. This was not possible for point-to-point variations which occur on a much smaller scale below the precision of the position reconstruction. The correction factor $C$ can be written as:
	\begin{equation}
	C=\frac{1}{1-P}\times \frac{1}{1+2\frac{d_{\bot}^2+d_{\bot}\times r-1/6\times r^2}{d^2}},	
	\end{equation}
	where the first factor is for the atmospheric extinction correction and the second factor the next order geometric correction. Here $P$ indicates the average extinction probability of a photon in direction of the considered telescope, which was obtained from Monte Carlo simulations as described in Section \ref{Sim}, $r$ the effective telescope radius (which is the same for all HESS-I telescopes) and $d_{\bot}$ the distance of the telescope mirror centre to the UAV in the direction perpendicular to $d$. As such, for each UAV-calibration event, the relative efficiencies $\epsilon_i$ of the different telescopes $i$ can be defined as:
	\begin{equation}
	    \epsilon_i=\frac{(I\times d^2\times C)_i}{\left\langle (I\times d^2 \times C)_j\right\rangle },	
	\end{equation}
	where $\left\langle (I\times d^2\times C)_j\right\rangle $ is the average of $I\times d^2\times C$ over all telescopes $j$ for the considered event. The run-wise relative efficiencies for each telescope were calculated simply by averaging over the efficiencies for the individual events. These relative efficiencies are the needed parameters to do the inter-calibration of the telescopes: Inter-calibrating the telescopes just means multiplying the intensity measured in each telescope by $1/\epsilon_i$.

	\subsection{Pointing Corrections}
	\label{PC}
	As described in Section \ref{UAVPos}, the position of the UAV was determined from the position of the UAV on the camera focal plane; specifically by minimising the sum of the squared distances of the UAV to the line of sight in which the UAV is for each telescope. The line of sight was obtained using the position of the image of the UAV in the camera defined by the centre of gravity of the image. Of course, the best fit position is not exactly on the determined lines of sight and so there are remaining residuals on the centre of gravity. These residuals are, among other, due to slight mispointings of the telescopes (mainly due to the weight and subsequent deformation of the telescope structure) leading to a wrong reconstructed direction. So, one can use these residuals to estimate the mispointings of the telescopes and possibly improve the pointing corrections used in Cherenkov telescopes if one is able to disentangle the shift of the residuals due to mispointings from the shift due to other physical phenomena.
	\par 
	To quantify the accuracy with which the UAV calibration events could identify mispointings, three different pointing models are compared: the so-called Null Model in which no pointing corrections are applied at all (i.e. one assumes that there is no structure deformation leading to mispointings and that the nominal pointing corresponds to the actual pointing) and two models obtained with the H.E.S.S. standard procedure for creating pointing correction models. One of them was constructed from data taken in November and December 2016 (the last one available at the moment of data taking) and the other one from data taken in May and June 2018, so covering the period where the UAV runs were carried out, except for CT4 where no data from this period is available due to a hardware failure (therefore the last available pointing model from December 2017 and January 2018 was used for CT4). This last pointing model has been used in this study, except for the part explicitly on the pointing corrections, as it is the one covering the period of data taking. In this standard procedure, a mechanical model of the telescope deformation as function of elevation and azimuth, that also includes the small tilt of the basement, is built. It has \SI{18}{} parameters and is constructed by pointing the H.E.S.S. telescopes with closed camera lid to bright stars selected in a way to get a nearly isotropic distribution across the visible sky. The light of these stars is reflected by the H.E.S.S. mirrors, then by the camera lid and finally recorded with a CCD camera in the centre of the dish where its position is compared with the position of light spots from LEDs in the camera frame \cite{Pointing}. From this the pointing deviation is computed and corrected for. 
	\par 
	Using the Monte Carlo simulations, which will be described in Section \ref{Sim}, 
	an estimation for the order of magnitude of the shift due to further physical effects beside mispointings is done. As broken pixels are expected to play a big role in these further effects, this estimation is done once with (using the actual broken pixels in the run on an event-by-event basis) and once without broken pixels. This allows us to give an estimation of how much one could improve the pointing corrections by being better able to recover the intensities of broken pixels. If the residuals are substantially larger with the three considered pointing models than in the simulation, it is being attempted to improve the pointing models by shifting the centre of gravity by its average offset and including this correction in the pointing model.


\section{Monte Carlo Simulations}
\label{Sim}
 To quantify the impact of aberration effects discussed in Section \ref{TC}, Monte Carlo simulations of a pulsed point-like calibration light source at a range of distances from the HESS-I telescopes were produced. In particular, these simulations assumed the light source to be a point source with isotropic emission (even though only photons which could potentially reach the telescopes were simulated) emitting pulses of a duration of \SI{4}{\nano\second} at a wavelength of \SI{400}{\nano\meter}. The photons were generated uniformly over the time interval defined by the pulse length. To increase computational efficiency of the simulation process, the quantum efficiency of the HESS-I photo-multiplier tubes was applied at source so as to not simulate and propagate photons which would not be detected. After the emission, the photons were propagated in straight lines until reaching the telescope. During this propagation, the photons which would have been extinct in the atmosphere were removed from the simulation and the integrated refractive index of the atmosphere was taken into account for computing the time of arrival of the photons at telescope altitude. This was done using the standard H.E.S.S. atmospheric model based on the considerations described in \cite{Bernloehr} using temperature and pressure profiles determined in balloon flights undertaken in 1999 in Windhoek as input parameters \cite{guy2003thesis}. Then, the photons were passed to the H.E.S.S. standard detector simulation which simulates the propagation of a photon from its position at telescope altitude via the mirror and camera to the pixels and then its conversion to photo-electrons to get the charge accumulated in each pixel\footnote{For more information about the H.E.S.S. detector simulations see for example \cite{guy2003thesis}.}. At the end, it simulates the trigger, amplification and the digitization of the signal, using realistic signal pulse shapes. For simulating the trigger, the whole charge accumulated in the camera during an event is considered. Then, the camera is divided in different partially overlapping sectors and the event is only kept if the following condition is fulfilled: at least three pixels in a sector exceed a charge threshold corresponding approximately to four photo-electrons \cite{hess-trigger}.
 \par
After this simulation procedure, the UAV-calibration events are in the same format as for the actual runs. To convert the charge measured by each pixel to photo-electrons, we followed the same procedure that was applied to the UAV-calibration events as discussed in Section \ref{analysis}. We do however note that, since the pedestal and gain calibration are relying on measurements, these also had to be simulated: for the pedestals \SI{50000}{} events with only night sky background were simulated and for the gains \SI{50000}{} events with the LED usually used for gain calibration. Finally, the same data cleaning and analysis procedure as described before for the taken data was applied to the simulated data.
 \par 
    \begin{figure}
		\centering
		\begin{subfigure}{.24\textwidth}
			\centering\includegraphics[width=1.\linewidth]{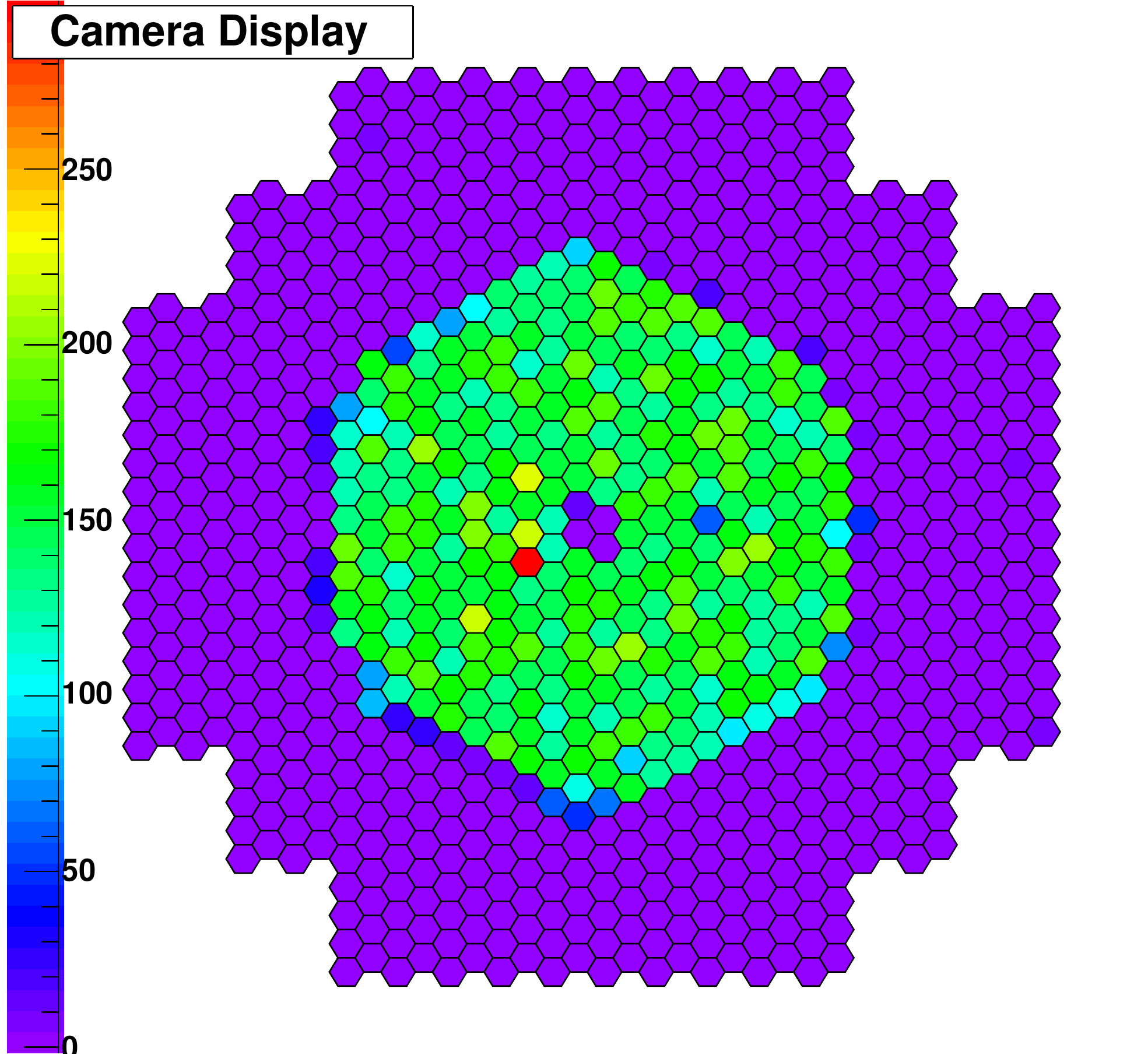}
		\end{subfigure} 
		\begin{subfigure}{.24\textwidth}
			\centering\includegraphics[width=1.\linewidth]{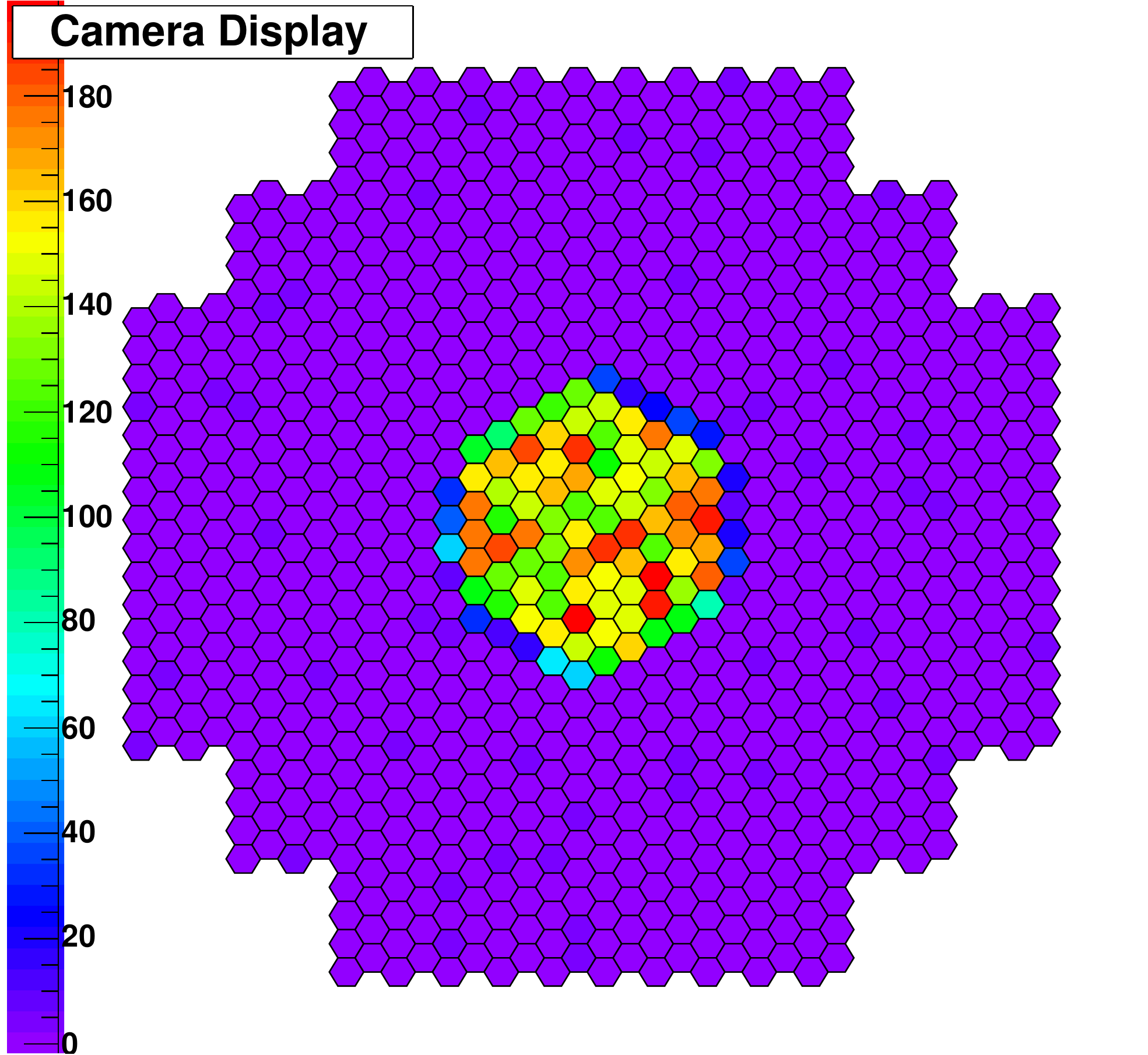}
		\end{subfigure}
		\begin{subfigure}{.24\textwidth}
			\centering\includegraphics[width=1.\linewidth]{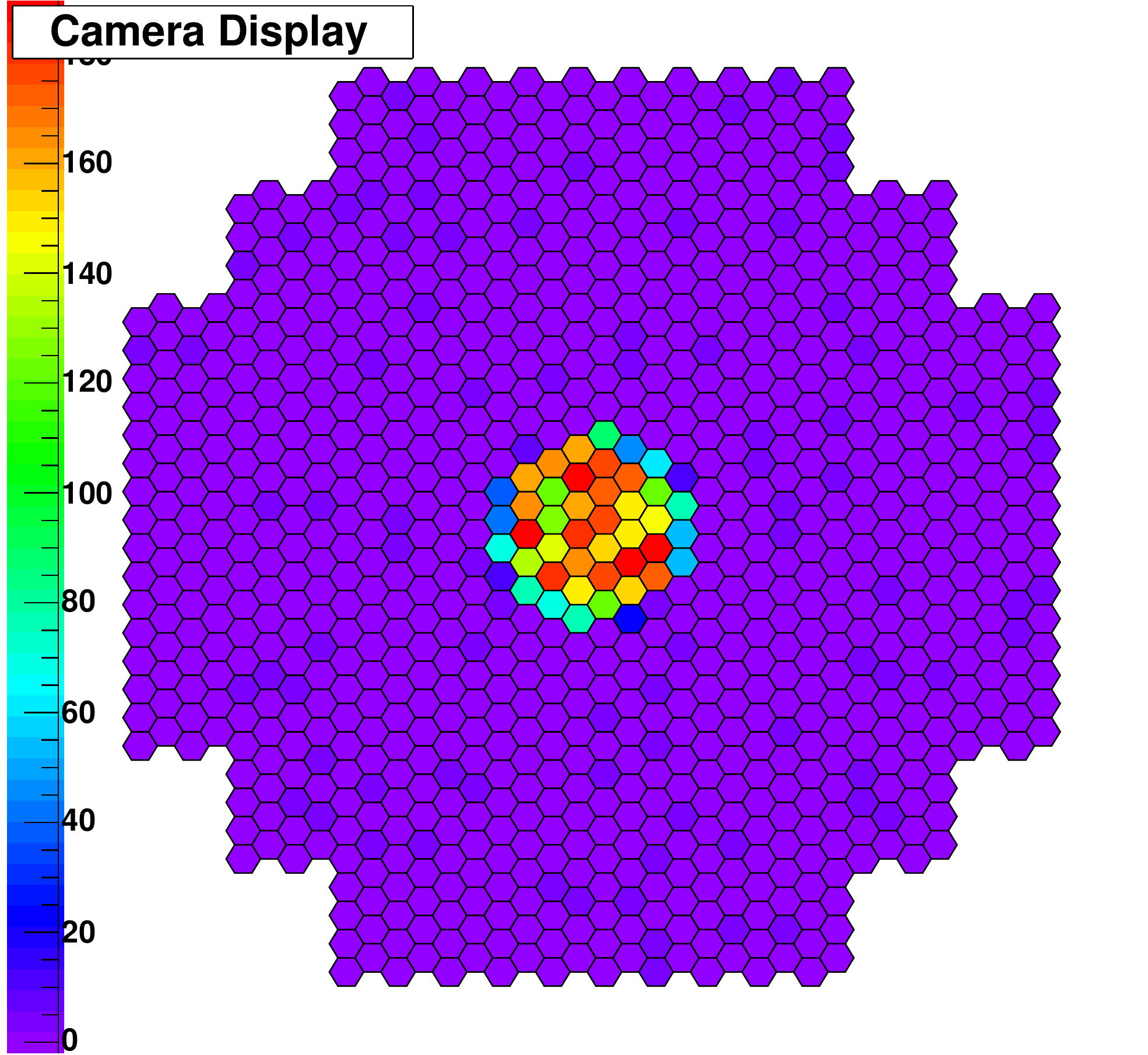}
		\end{subfigure} 
		\begin{subfigure}{.24\textwidth}
			\centering\includegraphics[width=1.\linewidth]{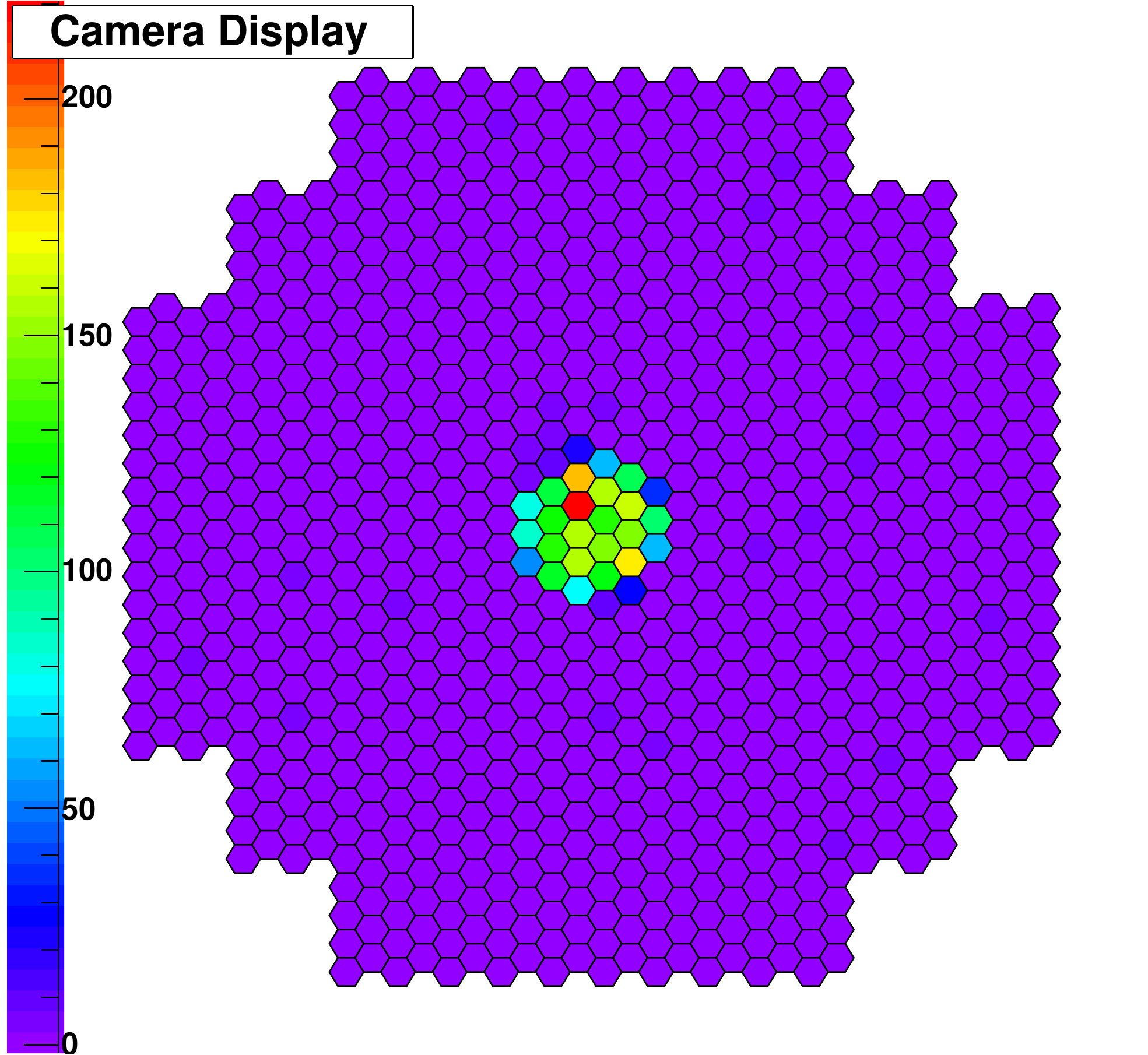}
		\end{subfigure} 
		\caption{Simulated events, as seen by HESS-I telescopes, of a calibration pulse emitted by a calibration system on a UAV holding a stationary position (a `hover') above the centre of the H.E.S.S. array at altitudes of \SI{250}{\meter}, \SI{500}{\meter}, \SI{750}{\meter} and \SI{1000}{\meter} (from left to right). The colour scale indicates the number of photo-electrons recorded in each pixel.}
		\label{ExSimEv}
	\end{figure}
 Examples of event displays resulting from the simulation showing the number of photo-electrons recorded in each camera pixel can be seen in Figure \ref{ExSimEv}, which clearly shows the effects of aberration and how the magnitude of these aberration effects is dependent on the separation between telescope and light source. 

	\begin{figure}
	    \begin{subfigure}{0.49\textwidth}
			\centering\includegraphics[width=1.\linewidth]{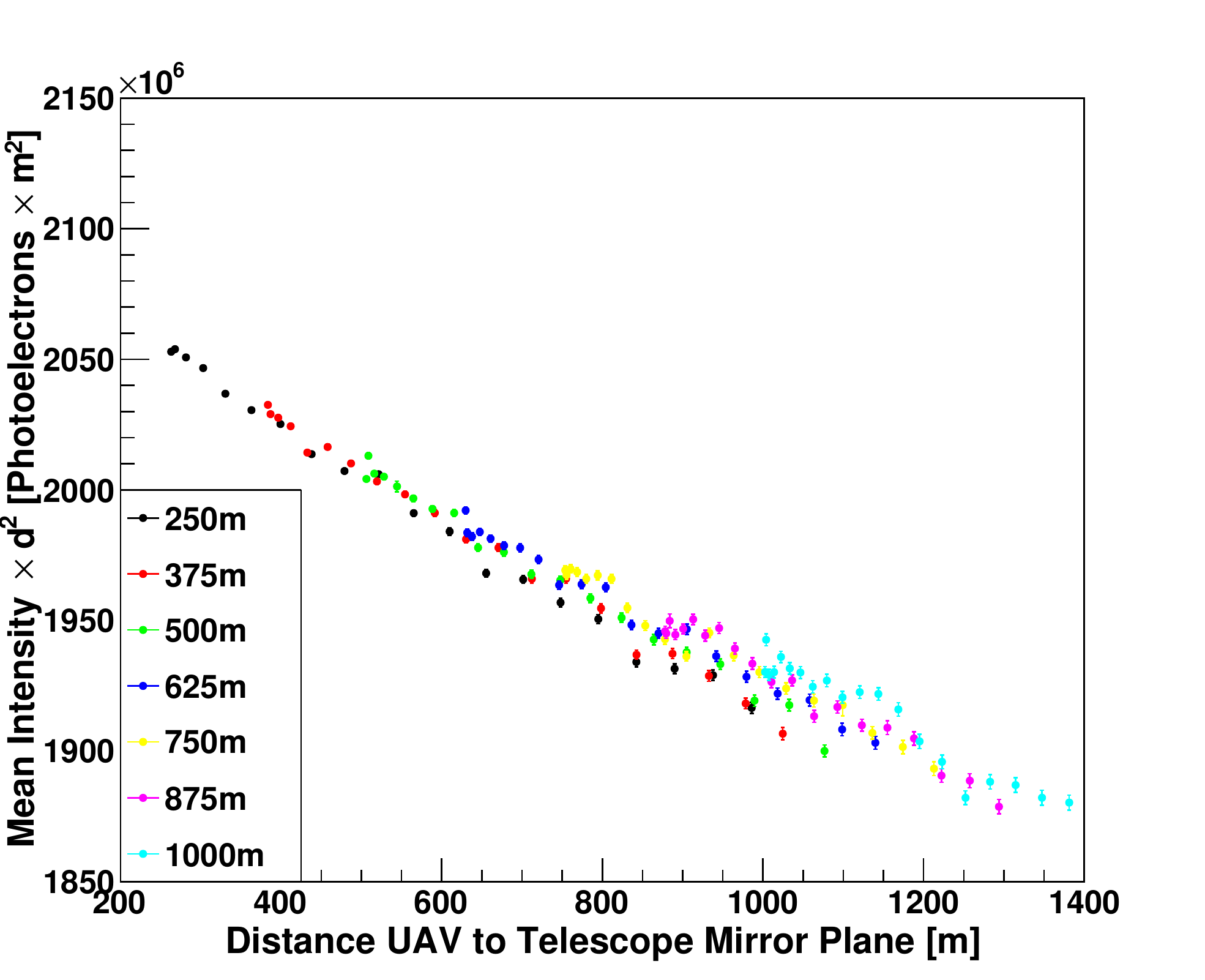}
		\end{subfigure} 
		\begin{subfigure}{0.49\textwidth}
			\centering\includegraphics[width=1.\linewidth]{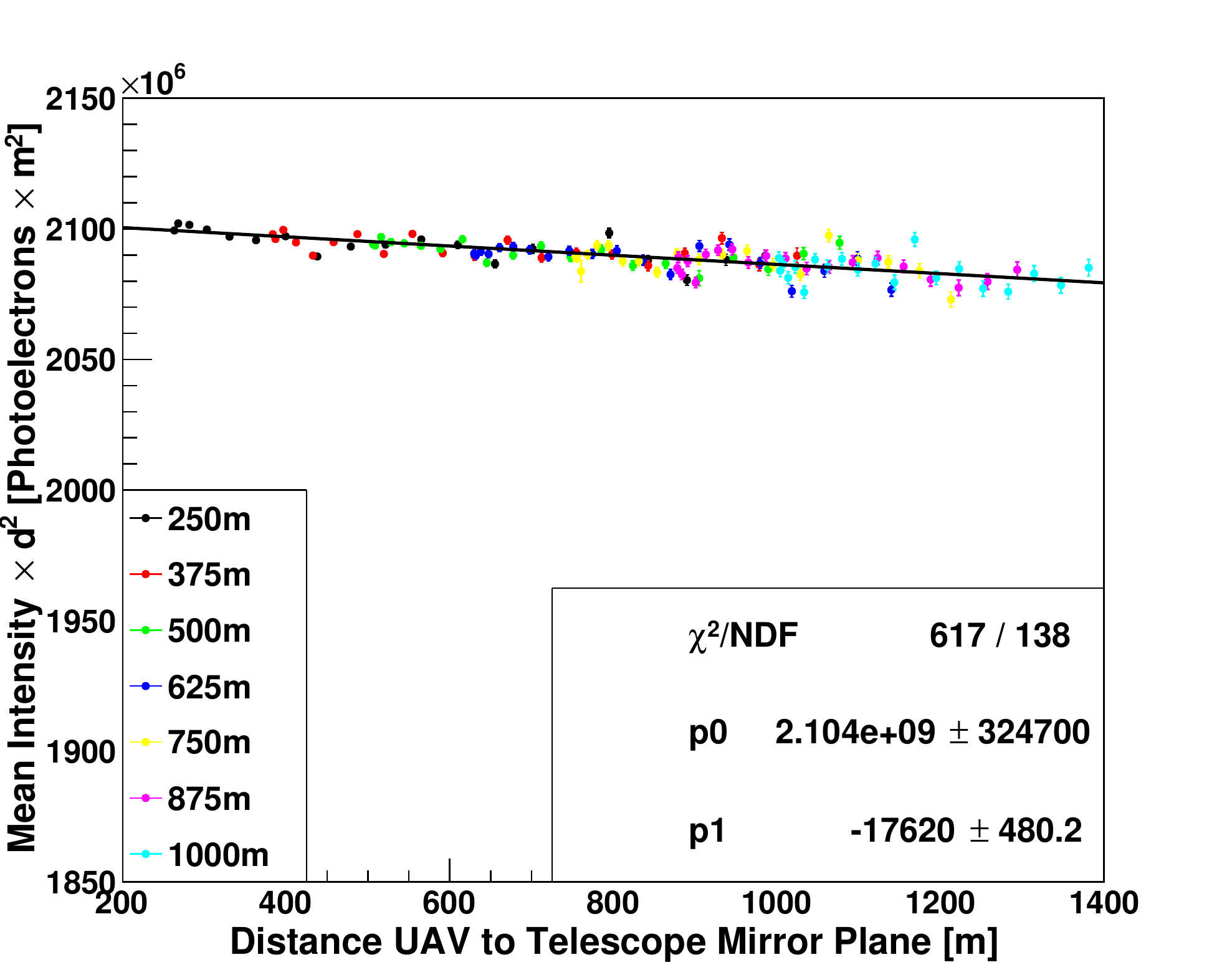}
		\end{subfigure} 
		\caption{$I\times d^2$ against $d$ for CT1 for events generated at 7 different altitudes and 20 different horizontal displacements of the UAV to the array centre. Left: Including atmospheric extinction in the simulation; Right: Simulation without atmospheric extinction}
		\label{int}
	\end{figure}

As discussed in Section \ref{inter-calibration} where the inter-calibration procedure is presented, the inter-calibration concept is based on $I\times d^2$ being constant against a change of $d$ (after a correction for the effects discussed in this paragraph). This is expected from geometrical considerations as the photons were emitted uniformly over solid angle\footnote{This is only approximately true: from geometrical considerations a shift of the UAV parallel to the mirror plane will also lead to a change of $I$ of the order of $\frac{1}{d^4}$ (instead of $\frac{1}{d^2}$ as for movement perpendicular to the mirror plane). This additional factor leads to a much smaller change than other neglected effects such as the point-to-point variations and is accounted for in the correction factor $C$.}. To check the validity of this expectation, additional simulations were conducted with the calibration light source at a large range of distances, both with and without atmospheric extinction of the light as it is propagated from calibration source to telescope. For the determination of the dependence of $I\times d^2$ on $d$, 500 calibration pulses, or events, were generated at different altitudes (\SI{7}{} regularly spaced altitudes between \SI{250}{\meter} and \SI{1000}{\meter}) and horizontal position offsets of the UAV to the centre of the array (\SI{20}{} regularly spaced offsets between \SI{0}{\meter} and \SI{950}{\meter}). The simulations were run here without any non-operational pixels, as their influence on the intensity is expected to depend on their distribution and can so not be accounted for by introducing a general correction factor, but have to be handled on a case-by-case basis respectively included in the systematic uncertainty, and without mispointings as they should be equivalent to a change of pointing direction of the telescopes. The resulting plot of $I\times d^2$ against $d$ is shown on the left of Figure \ref{int} for CT1. 
\par
$I\times d^2$ is slightly decreasing with $d$. This is due to the atmospheric extinction being neglected considering only geometrical arguments. Considering the atmospheric extinction too, one expects the more photons to be extinct, the more matter the photons pass (i.e. the higher the altitude of the UAV or the larger its horizontal offset to the centre of the array) leading to a behaviour similar as on the plot. To check whether the atmospheric extinction is responsible for the whole decrease, the simulations were repeated without atmospheric extinction. On the right plot of Figure \ref{int}, which shows the same plot as before without atmospheric extinction, only a decrease of about \SI{1}{\%} over the whole distance range is left and so almost the whole decrease was due to extinction. 
\par 
The extent of this remaining decrease depends strongly on the cleaning thresholds and completely disappears when reducing the camera signal thresholds applied during image cleaning, which indicates that a small fraction of the image is most likely cleaned away and that this fraction increases proportionally with the distance of the UAV to the telescopes (which reduces the abberation effects and thus the size of the UAV-calibration image is on the camera focal plane. Reducing the cleaning thresholds comes however at the expense of increasing the amount of accepted night sky background variations and so increasing statistical uncertainties. As this decrease is an order of magnitude smaller than the decrease due to the atmospheric extinction and also much smaller than the point-to-point variations discussed next, the cleaning thresholds were not decreased. As an additional check, the whole data analysis presented previously was also performed without cleaning, which only marginally impacted the inter-calibration results beside increasing statistical uncertainties.
\par 
However, even though there is almost no global variation with $d$ once atmospheric extinction is removed from the simulations, there are small point-to-point variations of about \SI{1}{\%} which are larger than the statistical uncertainties\footnote{ The statistical uncertainties were computed by taking $1/\sqrt{500}$ times the standard deviation of $I\times d^2$ for the $500$ events generated at each position.}. They are likely due to boundary effects when illuminating a different number of pixels and boundaries between pixels. This shows that $I\times d^2$ is constant against a change of $d$ modulo this \SI{1}{\%} point-to-point variations neglecting atmospheric extinction.
\par 
Including the atmospheric extinction again, the change of $I\times d^2$ over the relevant range of $d$, defined by the maximum distance between two telescopes of \SI{169.2}{\meter}, which is much bigger than the registered movement of the UAV in the considered data set as discussed in Section \ref{PosDetSection} (about \SI{30}{\meter} altitude variation), is also about \SI{1}{\%} for the given UAV telescope separation. Additional Monte Carlo simulations were performed, this time simulating the actual data taking runs with the UAV at its reconstructed position. From these simulations, the extinction probabilities of photons emitted in direction of the mirror centre of each individual telescope were computed for each recorded event and averaged. The average extinction probabilities for a photon in direction of the different telescopes were \SI{6.7}{\%}, \SI{7.6}{\%}, \SI{7.5}{\%} and \SI{6.6}{\%} for CT1-4 respectively. These extinction probabilities have been applied as correction factors while computing the relative efficiencies as discussed previously. They led to a change of the relative efficiencies of about \SI{0.5}{\%}.


\section{Results \& Discussion}
	\subsection{Inter-Calibration}
\begin{table}
	\centering
	\begin{tabular}{|c|c|c|c|c|c|c|}
		\hline
		Run&\multicolumn{2}{c|}{}&\multicolumn{2}{c|}{}&\multicolumn{2}{c|}{Muon}\\
		Identi-&\multicolumn{2}{c|}{A}&\multicolumn{2}{c|}{B}&\multicolumn{2}{c|}{(Observation}\\
		fication&\multicolumn{2}{c|}{}&\multicolumn{2}{c|}{}&\multicolumn{2}{c|}{Period Average)}\\
		\hline
		Tele-&\footnotesize{Relative}&\footnotesize{Statistical}&\footnotesize{Relative}&\footnotesize{Statistical}&\footnotesize{Relative}&\footnotesize{Statistical}\\
		scope&\footnotesize{Efficiency}&\footnotesize{Uncertainty}&\footnotesize{Efficiency}&\footnotesize{Uncertainty}&\footnotesize{Efficiency}&\footnotesize{Uncertainty}\\
		\hline
		1&0.929&0.001&0.942&0.001&0.9661&0.0002\\
		2&1.046&0.001&1.055&0.001&0.9872&0.0002\\
		3&1.097&0.002&1.073&0.001&1.0579&0.0002\\
		4&0.928&0.001&0.930&0.001&0.9889&0.0002\\
		\hline
	\end{tabular}
	\caption{Run-averaged relative efficiencies for each telescope in the two successful runs and their statistical uncertainties (standard uncertainty on the mean). The muon relative efficiency averaged over the observation period is given for comparison. In run A, $343$ four-telescope UAV events were recorded and in run B $350$.}
	\label{EfficienciesAbsApplied}
\end{table}	
\begin{table}
	\centering
	\begin{tabular}{|c|c|c|c|}
		\hline
		Run Identification&A&B&Muon\\
		\hline
		with respect to Default Calibration&8.5&7.5&4.0\\
		with respect to Run A&/&1.6&5.8\\
		with respect to Run B&1.6&/&5.4\\
		with respect to Muon Calibration&5.8&5.4&/\\
		\hline
	\end{tabular}
	\caption{Deviations between different relative calibrations in [\%] obtained for the different runs, the default inter-calibration (i.e. no correction for different efficiencies of the telescopes at all) and the previously used (period-averaged) muon inter-calibration.}
	\label{CalibrationDeviationsAbsApplied}
\end{table}

To derive the relative efficiencies for each telescope in the two UAV-calibration runs, we apply Eq. 2 to all events recorded in four telescopes. The resultant relative efficiencies and their statistical uncertainties are shown in Table \ref{EfficienciesAbsApplied}. To determine the deviations between the relative efficiencies obtained from the two UAV-calibration runs, we calculated the standard deviation of the UAV relative efficiencies by taking the sample standard deviation from \SI{0}{} of the differences of the UAV-derived relative efficiencies, with the sample standard deviation defined as:
	\begin{equation}
	    s = \sqrt{\frac{1}{N-1}\sum_{i=1}^N(x_i-\bar{x})^2}
	\label{sd}    
	\end{equation}
Here, $N$ is the number of used telescopes, i.e. $4$, $x_i$ the difference between the relative efficiencies of the compared runs for telescope $i$ and $\bar{x}$ the mean of the $x_i$'s which is $0$ by definition (as the relative efficiencies always sum up to $4$).
	\par 
Applying Equation \ref{sd} to the relative efficiencies derived from both UAV-calibration runs, we find that they deviate by \SI{1.6}{\%} as tabulated in Table \ref{CalibrationDeviationsAbsApplied}. This suggests that for these two runs, even though conducted on different nights, with different observing and environmental conditions, the UAV-based approach to inter-calibrating IACT arrays is able to deduce telescope relative efficiencies that are consistent within this margin, which, however, is much larger than the statistical uncertainties and constitutes a first evidence for the systematic uncertainties discussed later.
\par 
	\begin{figure}
	    \centering\includegraphics[width=13cm]{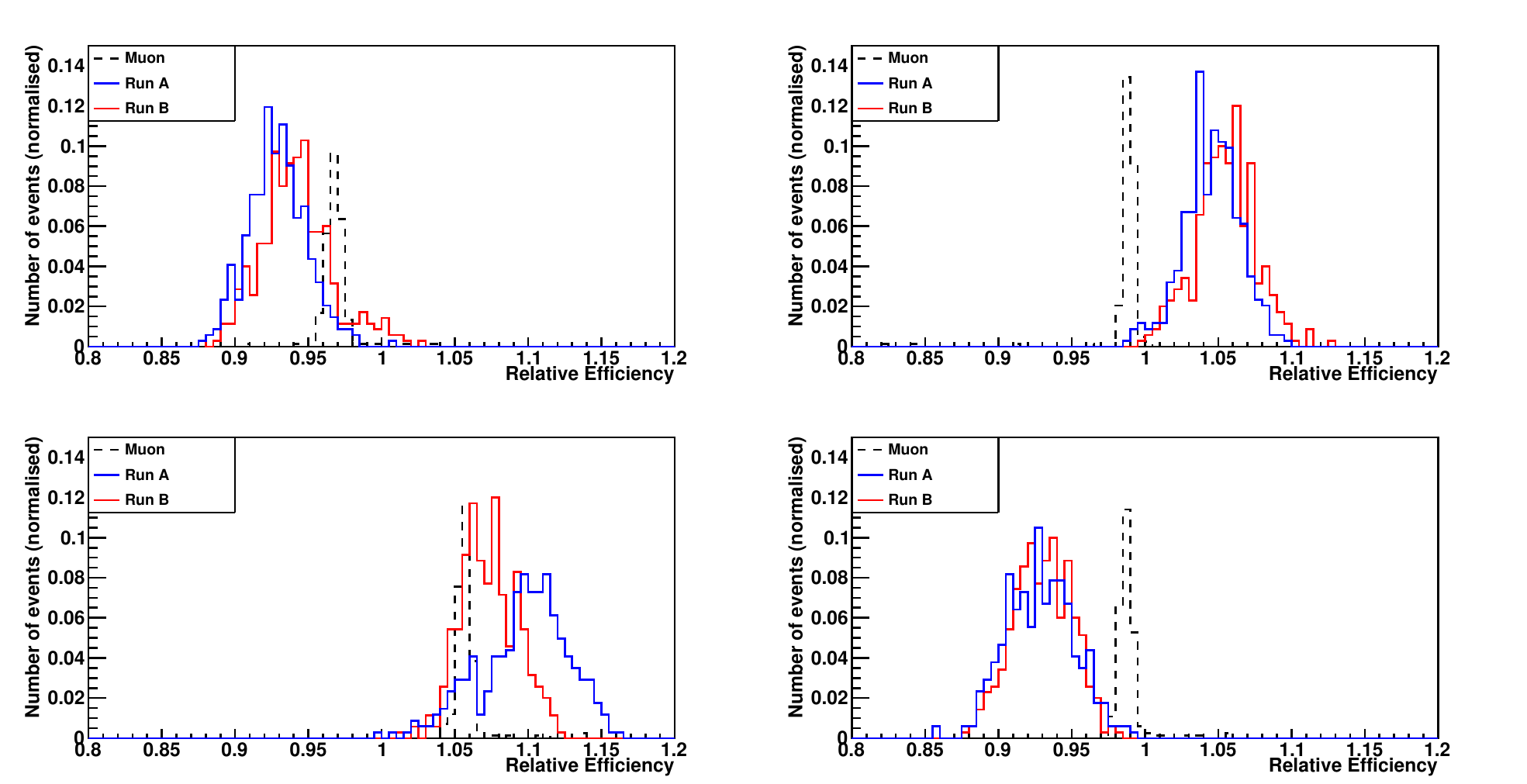}
		\caption{Normalised distribution of the relative efficiencies on an event-by-event basis as determined with the UAV for run A (blue) with 343 four-telescope UAV events and run B (red) with 350 four-telescope UAV events for all $4$ HESS-I telescopes (Top left: CT1, top right: CT2, bottom left: CT3, bottom right: CT4). In addition, the distribution of the relative muon efficiencies over the whole observation period on a run-by-run basis is shown in dashed black. It has been rescaled to the height of the other distributions for easy comparison.}
		\label{Eff}
	\end{figure}

The optical efficiency calibration of the H.E.S.S. telescopes has previously used the Cherenkov radiation from atmospheric muons as calibration light source~\cite{mitchell2016thesis}. Both the muon calibration and the UAV calibration are based on completely different processes and are so not expected to have any common systematic uncertainties beyond those associated with atmospheric propagation and the uncertainties intrinsic to any calibration procedure using a light source outside of the telescopes. The main common uncertainties are the telescope operational uncertainties present in all telescope observations, namely mostly the uncertainties on pedestal and flat-fielding. The cleaning could in principle also introduce an uncertainty, however as discussed in the Section 4, removing the cleaning leads to an increase in the statistical uncertainties while only marginally changing the inter-calibration results. Then there are the broken pixels which are interpolated and so might introduce an uncertainty for the UAV calibration. However, the muon calibration is based on the comparison of a recorded muon ring with a modelled muon ring on operational pixels as described in \cite{MuonHESS} and so is not impacted by this uncertainty. As such, the remaining primary sources of systematic uncertainty are the light source diffuser uniformity, the atmospheric extinction, the pedestals and the flat-fielding as common uncertainties. 
\par
As outlined in Section 2, the calibration light source contained a $50^{\circ}$ circular top-hat diffuser, specifically a Thorlabs ED1-C50. Whilst the `top-hat' optical design aims for isotropic illumination across the opening angle of the diffuser, independent studies have found a significant angle dependency for the illumination throughput of a single ED1-C50 diffuser \cite{diffuser}. However, we note that over the $\sim8^{\circ}$ angular extent of the H.E.S.S. array as seen by the UAV at distance of \SI{800}{\meter}, the diffuser intensity is reported to be uniform to within \SI{<5}{\%} \cite{diffuser}. Given the fact that the calibration light source used during the campaign consisted of four separate LED units illuminating four independent diffusers in random orientations, we expect a \SI{<2-3}{\%} variation in the intensity of the calibration light as seen by the individual telescopes due to the diffuser's performance. As such, the transmission properties of the calibration light source diffusers are potentially one of the dominant sources of systematic uncertainty of this first campaign, and will be addressed in a later campaign by evolving the current calibration light source design to include several holographic diffusers in series.
\par
As discussed in Section 4, the change in the derived relative efficiency values, due to assumption of average atmospheric conditions, was on the order of \SI{0.5}{\%}. It is difficult to qualify the uncertainty of this value, given the limited number of measurements available under Namibian atmospheric conditions and this so needs further investigations during which the UAV is moved to different positions. However we note that, as the muon optical efficiency calibration is based on data from multiple very different pointings (216 runs over 25 nights), the effect of the atmospheric extinction on the computed average relative efficiencies is expected to average out over these different pointings and as such, is not a common uncertainty for the relative efficiencies. 
\par
The uncertainty on the pedestals is mostly quantified by the pedestal width compared to which potential systematic offsets are completely negligible. The average high gain and low gain pedestal widths converted from ADC counts to intensities were between \SI{0.87}{} and \SI{1.09}{} photo-electrons and between \SI{0.76}{} and \SI{0.98}{} photo-electrons respectively. As the pedestal widths of the individual pixels are independent this number can be multiplied by the square-root of the average number of illuminated pixels which was between \SI{33}{} and \SI{47}{} to get the overall uncertainty due to pedestals in an event. This value was between \SI{4.6}{} and \SI{6.8} photo-electrons for the different telescopes and gain channels. These values need to be compared to the standard deviation of the recorded intensity in the UAV events which was between \SI{83}{} and \SI{176}{} photo-electrons and so more than an order of magnitude larger than the uncertainty due to pedestals. This shows that the statistical uncertainty in the recorded UAV light between the events is much more important than the statistical uncertainty due to pedestal width. 
\par
A similar comparison can be done for the flat-fielding uncertainty. The average flat-fielding uncertainty in each pixel is between \SI{0.4}{\%} and \SI{0.5}{\%} of the recorded intensity. Using that these relative uncertainties are statistically independent, this leads to an uncertainty between \SI{0.06}{\%} to \SI{0.08}{\%} over all the illuminated pixels. The relative uncertainty of the intensity of a single UAV event is between \SI{2}{\%} and \SI{4}{\%}. However, the effect of this uncertainty is statistically independent for all the UAV events whereas the effect of the uncertainty on the flat-fielding is the same for all the events as the flat-fielding is determined once for the whole observation period. For this reason, the relative uncertainty on the UAV intensity needs to be normalised to the number of events which leads to values between \SI{0.14}{\%} and \SI{0.22}{\%}. So also here the uncertainties due to the statistical variations in the intensity are higher, even though the difference between both uncertainties is not as large. 

As there are no other common uncertainties, the muon calibration method is well suited for a cross-check of the UAV inter-calibration method. For this reason, Table \ref{EfficienciesAbsApplied} also shows the relative efficiencies obtained with the standard muon calibration method, averaged over the whole observation period (25 nights). The averaging over the observation period is necessary due to the run-by-run variations of the muon optical efficiencies and to reduce the impact of the uncertainty in the atmospheric extinction model. The relative efficiencies obtained from the UAV calibration deviate by \SI{5.8}{\%} and \SI{5.4}{\%} from the relative efficiencies obtained from the muon calibration for run A and B respectively as shown in Table \ref{CalibrationDeviationsAbsApplied}. As both methods are not expected to have any systematics in common beside telescope operational uncertainties, this consistency within about \SI{5.5}{\%} for the different runs, is an indication that the uncertainties of both methods are of the same order of magnitude or less.

The normalised distributions of the UAV-derived relative efficiencies, along with the muon-derived relative efficiencies are shown in Figure \ref{Eff}, for CT1 through CT4 respectively. For the UAV distributions, we show the efficiencies on an event-by-event basis for both UAV runs normalised to the total number of events, while for the muon distribution we show the efficiencies on a run-by-run basis, normalised to the total number of muon runs considered. As can be seen in Figure \ref{Eff}, the width of the distributions for the UAV-derived relative efficiencies are similar, as can also be seen in the uncertainties stated in Table \ref{EfficienciesAbsApplied}, though for some telescopes there appears to be an offset between the means of the distributions for the different runs. The similar size of the distribution widths implies that, even with our first-generation UAV-calibration system, we are able to reproduce the statistical fluctuation of the technique from one UAV-calibration run to the next. The offset in the means of the distributions implies a systematic effect, for example, due to differences in the atmospheric conditions at the time of the individual UAV-calibration runs. Determining the source of this offset requires additional UAV-calibration runs in a variety of atmospheric conditions and UAV positions. As such, this goes beyond the scope of this paper and will be addressed in future UAV-calibration campaigns at the H.E.S.S. array. 

\par 
\begin{figure}
	\centering\includegraphics[width=12cm]{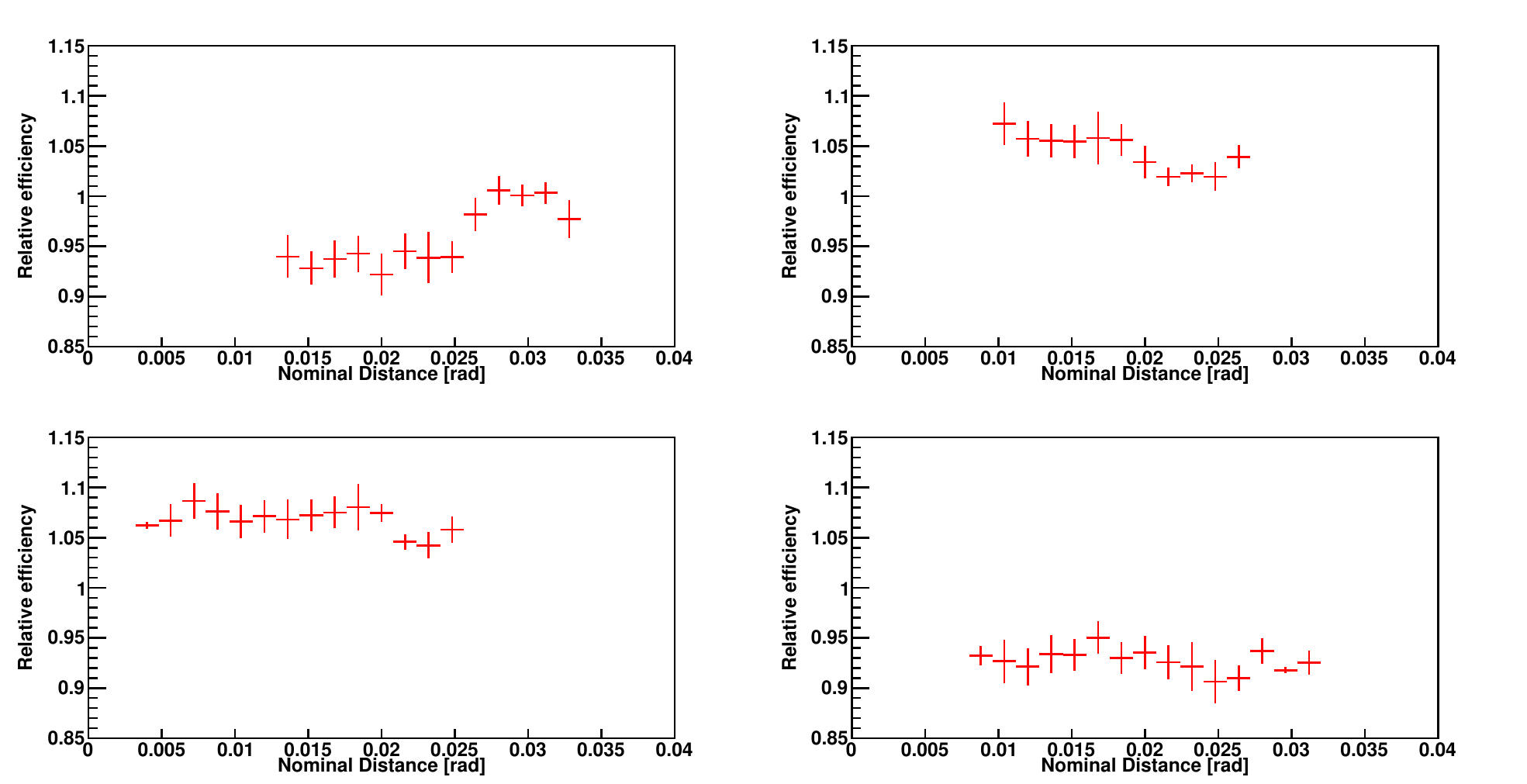}
	\caption{Relative efficiency for all $4$ HESS-I telescopes as a function of radial distance from the centre of the camera for all 4-telescope events in UAV-calibration run B. The data were binned into \SI{6.25}{\milli\radian} radial bins, with the uncertainty bars indicating the standard deviation of each bin value.}
	\label{138951RelEffND}
\end{figure}

Throughout calibration run B, the UAV periodically moved in and out of the telescopes' field of view. By binning the UAV-calibration events in run B as a function of radial distance from the centre of the camera to the centre of gravity of the UAV image, we are able to investigate the radial dependence of the computed relative efficiency. This is shown in Figure \ref{138951RelEffND} where all events in run B were binned into \SI{6.25}{\milli\radian} radial bins, with the uncertainty bars indicating the standard deviation of each bin value. The radial distributions of Figure \ref{138951RelEffND} indicate that for CT1-4 the relative efficiencies vary across the camera's field of view. The magnitude of this variation is telescope dependent.
\par 
This difference between relative efficiencies of run A and run B, as seen in Figure \ref{Eff} and the radial dependence of the relative efficiency for run B, as seen in Figure \ref{138951RelEffND} can be due to a number of phenomena. First of all, the broken pixels lead to missing intensity in the image. Even though they are interpolated, it is not possible to exactly recover the amount of light which hit them, leading to a change of the total intensity highly dependent on the exact position of the image in the camera. Second, systematic uncertainties in the position determination lead to an imprecise correction of the expected intensity for the distance of the UAV to the telescope mirror, which could depend on position and which alters the computed relative efficiencies. Third, the point-to-point variations found in the simulation which have been neglected play a role at percent level. Fourth, the uncertainty in the atmospheric extinction model and possible differences in the atmospheric conditions between the two runs might play a role. And last but not least, the uncertainties in the flat-fielding, the difference between the wavelengths of the flat-fielding LEDs (\SI{370}{\nano\meter}) and of the UAV-mounted light source (\SI{400}{\nano\meter}) and a possibly inhomogeneous mirror response could also introduce a dependence on the camera position of the image on the found relative efficiencies. Investigating these factors in depth requires more UAV-calibration runs to be performed and as such, is beyond the scope of this paper. 

	\subsection{Pointing Corrections}
	\begin{figure}
		\centering
		\begin{subfigure}{1.\textwidth}
			\centering\includegraphics[width=1.\linewidth]{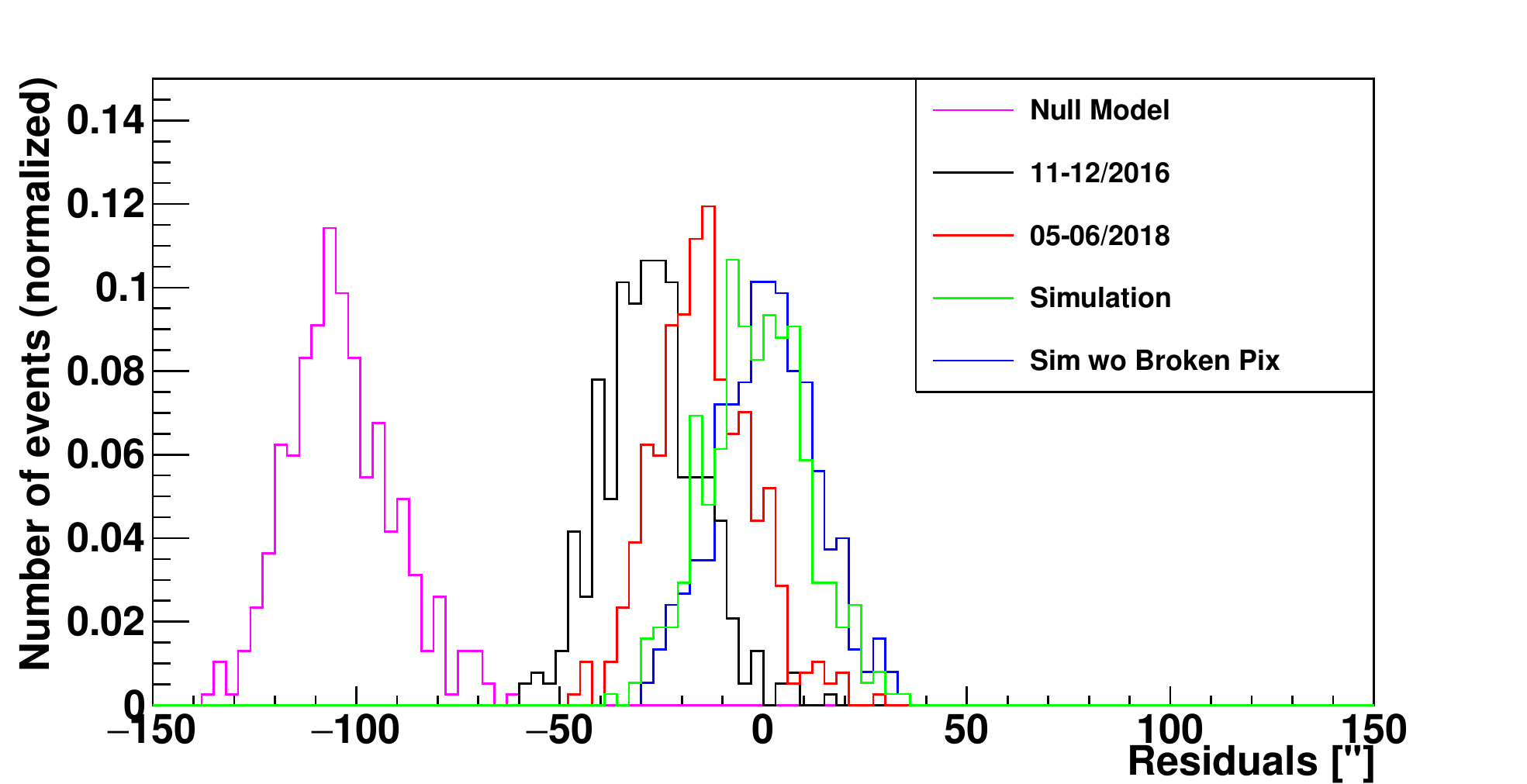}
		\end{subfigure} 
		\begin{subfigure}{1.\textwidth}
			\centering\includegraphics[width=1.\linewidth]{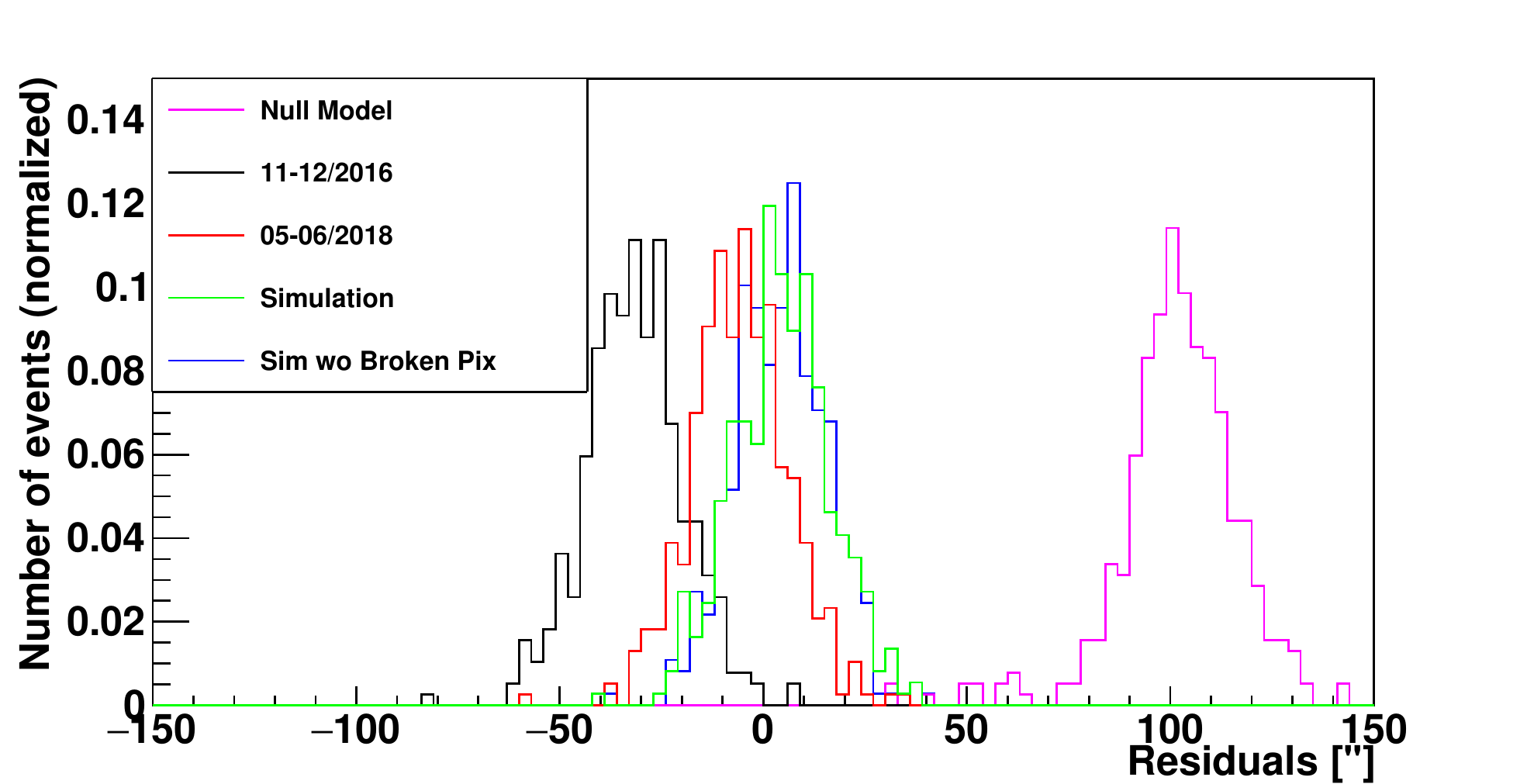}
		\end{subfigure}
		\caption{Distribution of residuals on centre of gravity for different pointing models described in Section \ref{PC} for one telescope and one of the two camera coordinates per run. The shown telescope and camera coordinate were chosen in a way that the residual distribution corresponds the most to the average distribution for the model and run (as they were looking quite differently for the different telescopes and runs). Top: Residuals on x-coordinate of centre of gravity in CT2 for run A. Bottom: Residuals on x-coordinate of centre of gravity in CT3 for run B.}
		\label{PointMod}
	\end{figure}
	\begin{table}
		\centering
		\begin{tabular}{|c|c|c|c|c|}
			\hline 
			Run&\multicolumn{2}{c|}{A}&\multicolumn{2}{c|}{B}\\
			Identification&\multicolumn{2}{c|}{}&\multicolumn{2}{c|}{}\\
			\hline
			Pointing&Quad Mean&Quad Mean&Quad Mean&Quad Mean\\
			Model&Residual&Spread&Residual&Spread\\
			\hline
			Null Model&60.24&17.11&63.94&22.80\\
			11-12/2016&35.92&16.37&34.38&17.15\\
			05-06/2018&9.31&15.05&5.77&15.99\\
			Simulation&6.61&12.05&5.84&12.65\\
			Sim wo Broken Pix&3.44&11.37&2.96&10.92\\
			\hline
		\end{tabular}
		\caption{Quadratic mean (over telescopes and position coordinates) of the average residuals and of the spread of the residuals (in arc-seconds) for the two runs and the different pointing models. Null Model: Model without any pointing corrections; 11-12/2016: H.E.S.S. standard pointing model based on data from November and December 2016; 05-06/2018: H.E.S.S. standard pointing model based on data from May and June 2018 (i.e. taken around the measurement period), except for CT4 where data from December 2017 and January 2018 was used; Simulation: Residuals obtained from simulation using broken pixels detected in data of runs and perfect pointing; Sim wo Broken Pix: Residuals obtained from simulation without broken pixels and perfect pointing}
		\label{PointRes}
	\end{table}
	As discussed in Section \ref{PC}, we compared the residuals of the position determination on the centre of gravity of the calibration image on the camera focal plane for three telescope pointing models and two simulation models (with and without broken pixels) to quantify the pointing accuracy of the HESS-I telescopes with the UAV-calibration runs. Distributions of these residuals for the five cases are shown in Figure \ref{PointMod} for illustration purposes. Additionally, the quadratic means (i.e. the root mean square) over the four HESS-I telescopes and over the x-y camera coordinates of the residuals and of the spread of the residuals are tabulated in Table \ref{PointRes} for each case and run. Comparing first only the three pointing models used on the taken data, the residuals from the Null Model are much higher than the residuals using pointing corrections (factor \SI{6}{} to \SI{11} using the most recent pointing model). This shows that using pointing corrections, we are better able to determine the position of the UAV and so that the pointings are more consistent. Thus, the UAV data provide an additional way to show that the pointing corrections are working and improving the knowledge of the pointing direction of the H.E.S.S. telescopes.
	\par 
	Comparing the pointing models from late 2016 to the ones covering the UAV-calibration campaign in May 2018, we find about a factor $5$ difference in the quadratic means of the residual distributions, \SI{9.31}{\arcsecond} and \SI{5.77}{\arcsecond} respectively for the 2018 pointing model compared to \SI{35.92}{\arcsecond} and \SI{34.38}{\arcsecond} respectively  for the 2016 pointing model. This demonstrates again the importance of using up-to-date pointing models as the absolute pointing of the telescopes changes over time due to processes such as the settling of the telescope foundations, or aging of the telescope structure. As can be seen in Figure \ref{PointMod}, using out-dated pointing models results in a larger uncertainty in the pointing direction of the telescopes, rendering the pointing corrections partially ineffective.
	\par 
	In general, Figure \ref{PointMod} highlights the possibility that UAV-calibration events can be used to investigate the accuracy of a telescope's pointing model, but this leaves the question open whether it is possible to improve the pointing corrections with UAV-calibration events. For this reason, simulations using perfect pointing and including all the other physical phenomena as much as possible\footnote{Indeed, as discussed in Section \ref{Sim}, the Monte Carlo simulation accounts for the atmospheric extinction and the refractive index of the atmosphere. In addition, the simulation contains a model of the H.E.S.S telescopes with the actual position of the photomultiplier tubes, drawers and the mirror facets. Photons are so fully propagated up to the facets level and so the effect due to segmented mirrors and optical aberrations are included in the simulation. The radial-offset-dependent PSF of the individual facets is set in a way that the simulated PSF matches the measured PSF and the shadow of the camera, the mast and other part of the telescope structure is taken into account by a general correction factor. Then, the triggering, readout and calibration is simulated as described in Section \ref{Sim}. More detailed information about what is included in the simulation can be found in \cite{guy2003thesis}} were run to disentangle the part of the residuals due to mispointings from the part due to other camera operational considerations. In particular, we ran two sets of Monte Carlo simulations, one using the broken pixels as determined in the simulated calibration run, and one with all pixels in the HESS-I cameras operational. Figure \ref{PointMod} and Table \ref{PointRes} show that the quadratic mean of the residuals is about $2$ times higher in the simulation with broken pixels (which have been interpolated during the data analysis as described in Section \ref{TC}) than in the one without broken pixels. The interpolated broken pixels also lead to a shift of the centre of gravity of the image even though the shift due to mispointings is much larger (as can be seen by comparing the Null Model to data driven pointing models). As such, considering all distributions in Figure \ref{PointMod} we can see that, whilst the telescope pointing appears to be the dominant contributor to shifting the residual distributions away from 0, other factors, such as broken camera pixels, also play a role and as such, it is not possible to completely remove the residuals in the UAV-calibration runs presented here, by simply improving the telescope pointing model. 
	\par 
	Comparing the residuals obtained from the simulation with broken pixels to the residuals of the UAV-calibration events obtained using the standard H.E.S.S. pointing corrections, we find that while the residuals of the broken pixel simulation are slightly smaller for run A and slightly larger for run B, overall there is no significant difference. The similar size of the residuals for these simulated and real calibration events indicates that the standard pointing corrections already reach the maximum precision achievable with the UAV without a more elaborated method to recover broken pixels. For this reason, a possible method for improving the residuals was investigated using the pointing model from 2016. The coordinates of the centre of gravity in each telescope were shifted by their determined average (over all the UAV events of a given run) offset. The obtained quadratic mean of the residuals was \SI{1.30}{\arcsecond} and \SI{1.39}{\arcsecond} for the two different runs. Applying the method iteratively led to smaller residuals (about \SI{0.02}{\arcsecond} for three iterations for example). This shows that shifting the coordinates of the centre of gravity by their average offset is a very efficient method to lower the residuals. It should however, be noted that this approach shifted the mean of the residual distributions to be lower than that of the simulation (in which perfect pointing was assumed). This suggests that this approach ``over-corrects'' the data. The shifting of the centre of gravity lowers the residuals no matter where they come from, i.e., the shift does not only account for mispointings, but also for broken pixels and other effects such as optical aberrations which are not necessarily consistent over multiple runs. One would have to disentangle the shift in residuals due to mispointings and the shift due to other effects. This could partially be done by increasing the number of configurations in which UAV-calibration data is taken (different positions of the UAV, trying to illuminate the telescopes evenly over the whole field of view) which would allow to eliminate effects due to the position of the image of the UAV on the camera such as illuminating always the same broken pixels. Additionally, one could go further in trying to recover the light in broken pixels: Instead of interpolating taking the average of the six neighbouring pixels, one could recover its intensity by fitting a model image to the recorded data. Last but not least, it might be possible to get the size of the shift from simulation and so not to take into account the part of the shift present in the simulation too for the correction.
	\par  
	A previous study found that the uncertainty of source position determination with H.E.S.S. due to systematic pointing uncertainties was between \SI{10}{\arcsecond} and \SI{20}{\arcsecond} per axis \cite{Gillessen}, which results in a pointing uncertainty of \SI{20}{\arcsecond} to \SI{40}{\arcsecond} per axis per telescope assuming the mispointings of the four telescopes to be independent. This has been independently confirmed by observing stars with known positions passing through the field of view of a H.E.S.S. telescope. In particular, the distributions of the angular distances between the measured and observed positions of these passing stars was found to be Gaussian with a standard deviation of \SI{15}{\arcsecond} to \SI{20}{\arcsecond} (per axis) depending on the exact configuration. The residuals obtained with the most recent pointing model are slightly smaller (Table \ref{PointRes}) showing that  the UAV already now achieves similar accuracy without the need for more elaborate data cleaning methods to recover broken pixels and that the standard quoted pointing uncertainties might be a bit overestimated. Another potential explanation could be that there are systematic mispointings consistent between the telescopes which cannot be detected looking at the residuals of the centre of gravity.
	
	
\section{Conclusions}
In this paper we present the results of the first ever inter-calibration of a Cherenkov telescope array with a UAV-based calibration light source. This UAV inter-calibration data was generated using a first-generation prototype consisting of a bespoke LED-based light source, emitting 4~ns long pulses of 400~nm light, housed on an off-the-shelf multi-rotor UAV. The UAV system was flown into the field of view of the four HESS-I telescopes of the H.E.S.S. array, resulting in the calibration pulses being recorded. The HESS-I telescopes were then inter-calibrated based on the total amount of light recorded in the different cameras. The obtained inter-calibration was consistent within \SI{5.8}{\%} and \SI{5.4}{\%} respectively with the muon inter-calibration for both of the runs. As both of these inter-calibration methods are based on very different physical processes, they are not expected to have any common systematic uncertainties, beside those intrinsic to all calibration methods based on a light source at a distance hundreds of meters from the telescopes. As these systematic uncertainties intrinsic to all calibration methods based on a light source at a distance hundreds of meters from the telescopes are on a smaller scale as discussed previously, this is an indication for both methods having uncertainties of this order of magnitude or less. This would mean that UAVs are well suited to inter-calibrate Cherenkov telescope arrays and that inter-calibrations with a single light source on an event-by-event basis would indeed be possible. 
\par
Importantly, this result indicates that UAV-based inter-calibration already delivers results with uncertainties at few percentage level at its first attempt with a non-optimised first-generation UAV prototype. This uncertainty will be improved through a better understanding of the systematic uncertainties of the technique by including more physical phenomena in their determination and by comparing the results of the UAV calibration to further independent methods beyond the muon-based calibration such as the air shower optical efficiency calibration method \cite{CrossCalibrationShowers}, as well as further iterations of the UAV prototype with a bespoke UAV platform, improvements to the calibration light source and a better integration of the calibration payload to the flight platform~\cite{2018Brown}. 
\par 
Beyond inter-calibration, we have also shown that, without taking any additional data, a UAV-based calibration light source also allows us to verify the pointing corrections of the H.E.S.S. telescopes by comparing the effect that different telescope pointing models have on the observed data. Indeed, we demonstrated that it is an additional method to show that first the pointing corrections of H.E.S.S. improve the direction reconstruction of incident Cherenkov photons with respect to using no pointing corrections at all and second that it is very important to use a recent pointing model due to a change of the pointing of the telescopes with time (among other due to the sinking of the foundation of the telescopes into the ground) leading to outdated pointing corrections which become ineffective. In its final implementation, this would of course not only be done for one given configuration, but would have to be repeated with the UAV at numerous different positions to verify a complete pointing model and not only verify it locally.
\par
Finally we note that, beyond inter-calibrating the relative efficiencies of Cherenkov telescopes and confirming their pointing accuracy, the flexibility and versatility of a UAV-based calibration system allow us to address other key calibration issues. For example, with the UAV-based system it is -- unlike muon inter-calibration -- possible to perform a multi-wavelength inter-calibration: one can just mount a different coloured calibration light source. This will allow us to monitor wavelength dependent effects, such as the wavelength dependent degradation of the telescopes' optical system and the wavelength dependency of the quantum efficiency of the photo-multiplier tubes. Finally we note that a UAV-based system will allow us to monitor the transparency of the lowest layers of the atmosphere with the UAV, either by mounting meteorological instruments on the UAV (as proposed in \cite{2018Brown}) or by trying to infer the atmospheric extinction from the amount of light recorded in the different telescopes. Future planned UAV-calibration campaigns will not only allow to build upon the success of this first UAV-based calibration campaign, it will also allow us to quantify the potential of a UAV system for these calibration requirements. 

	\section*{Acknowledgements}
	We thank Prof. S. Wagner, director of the H.E.S.S. Collaboration and Prof. O. Reimer, chairman of the H.E.S.S. Collaboration board, for allowing us to use H.E.S.S. data in this publication. AMB acknowledges the financial support of the Royal Society Research Grant RG160883 that funded the construction of the prototype used in this campaign, and funded the logistics associated with the campaign. We would also like to thank Matthias Büchele and David Jankowsky, the on-site shiftcrew during the UAV campaign, for their help in setting up the H.E.S.S. array without whom it would not have been possible to take any data. Alison Mitchell and Vincent Marandon for their help with the muon calibration. And Jean-Philippe Lenain for helping us to understand the H.E.S.S. detector simulation and applying it to the UAV data. Finally, our thanks go to all the members of the H.E.S.S. Collaboration for their technical support and for many stimulating discussions which undoubtedly improved the quality of the paper. 

	\bibliographystyle{elsarticle-num}
	\addcontentsline{toc}{part}{Bibliography}
	\bibliography{additional}
	
\end{document}